\documentclass[useAMS,usegraphicx,usenatbib]{mn2e}

\title[Evidence for a mixture of electron/positron pairs in AGN jets]
{Evidence for a significant mixture of electron/positron pairs in FRII jets constrained by cocoon dynamics}
\author[N. Kawakatu et al.]{Nozomu Kawakatu$^{1}$
\thanks{E-mail: kawakatsu@kure-nct.ac.jp (NK)}, 
Motoki Kino $^{2}$ and Fumio Takahara $^{3}$ \\
$^{1}$ Faculty of Natural Sciences, 
National Institute of Technology, Kure College, 2-2-11 Agaminami, 
Kure, Hiroshima, 737-8506, Japan \\
$^{2}$ Korea Astronomy and Space Science Institute,
776 Daedeokdae-ro, Yuseong-gu, Daejeon 305-348, Republic of Korea \\
$^{3}$ Department of Earth and Space Science, Osaka University, 
560-0043 Toyonaka, Japan}

\begin{document}

\date{}

\pagerange{\pageref{firstpage}--\pageref{lastpage}} \pubyear{2015}
\twocolumn

\maketitle

\label{firstpage}

\begin{abstract}
We examine the plasma composition of relativistic jets in four FRII radio galaxies by analyzing the total cocoon pressure 
in terms of partial pressures of thermal and non-thermal electrons/positrons and protons. The total cocoon pressure is 
determined by cocoon dynamics via comparison of theoretical model with the observed cocoon shape. 
By inserting the observed number density of non-thermal electrons/positrons and the upper limit of thermal 
electron/positron number density into the equation of state, the number density of protons is constrained.  
We apply this method to four FRII radio galaxies (Cygnus$\,$A, 3C$\,$219, 3C$\,$223 and 3C$\,$284), for which 
the total cocoon pressures have been already evaluated.   We find that the positron-free plasma comprising of 
protons and electrons is  ruled out, when we consider plausible particle distribution functions. 
In other words, the mixture of positrons is required for all four FRII radio galaxies;  the number density ratio of 
electrons/positrons to protons is larger than two. 
Thus, we find that the plasma composition is independent of the jet power and the size of cocoons. 
We also investigate the additional contribution of thermal electrons/positrons and protons on the cocoon dynamics. 
When thermal electrons/positrons are absent, the cocoon is supported by the electron/proton plasma pressure, 
while both electron/positron pressure supported and electron/proton plasma pressure supported cocoons are allowed 
if the number density of thermal electrons/positrons is about $10$  times larger than that of non-thermal ones. 
\end{abstract}

\section{Introduction}
Formation mechanism of relativistic jets in active galactic nuclei (AGNs) and its 
connection to the accretion disk  remain as an open issue in astrophysics 
(e.g., Rees 1984; Meier 2003; McKinney 2006; Komissarov et al. 2007; 
Tchekhovskoy et al. 2011; McKinney et al. 2012; Toma \& Takahara 2013; 
Nakamura and Asada 2013; McKinney et al. 2014; Ghisellini et a. 2014). 
The plasma composition is one of the issues that impede our 
understanding of AGN jets (e.g., Begelman et al. 1984 for review), 
because it is difficult to explore properties of bulk population such as low energy 
electrons/positrons and protons, whose emission timescale is too long and 
the relevant frequency is far below the observable range. 
Since the total kinetic power of AGNs is proportional to the mass density 
of jets, the power depends on the composition of outflowing matter. 
Thus, an uncertainty in matter contents results in an uncertainty of the total 
power, which is one of the basic physical quantities describing the jets. 
The evaluation of the actual kinetic power of jets is essential to understand 
the energetics and evolution of radio lobes (e.g., Falle 1991; Bicknell 1994; 
Kaiser \& Alexander 1997; Kino \& Kawakatu 2005; Ito et al. 2008; 
Kawakatu et al. 2008; Nakamura et al. 2008; Godfrey \& Shabala 2013; 
Maciel  \& Alexander 2014). 
Also, the composition is closely linked to the formation mechanism 
of relativistic jets since the acceleration and deceleration of jets near supermassive 
black holes strongly depend on the plasma compositions (e.g., Blandford \& Znajek 1977; 
Blandford \& Payne 1982; Phinney 1982; Konigl 1989; Sikora 1996; 
Inoue \& Takahara 1996). 

Up to now, the plasma composition of AGN jets, 
especially the fraction of kinetic energy stored in protons and cold leptons, 
has been investigated by a large number of papers. 
Because of the low radiative efficiencies, it is 
difficult to detect electromagnetic signals from these particles. 
To constrain the plasma compositions of discrete 
blobs in blazar jets, a variety of approaches have been proposed, on the basis of 
the synchrotron self-absorption for M87, 3C279 and 3C 345 (e.g., Reynolds et al. 1996; 
Hirotani et al. 1999, 2000; Hirotani 2005) and on the basis of the detection of circular polarization 
of the radio emission in 3C 279 jets (Wardle et al. 1998; Homan 2009). 
These indirect arguments have suggested the existence of electron/positron  
plasma in jets (see, however, Celotti \& Fabian 1993; Ruszkowski \& Begelman 2002).
Another approach to estimating the electron/positron pair plasma 
composition of blazar jets is the detection of bulk-Compton emission, which was proposed by 
Begelman \& Sikora (1987) [see Sikora \& Madejski (2000); Moderski et al. 2004; Celotti, 
Ghisellini \& Fabian 2007]. This method is applied to the blazar-scale knots (Kataoka et al. 2008; 
Watanabe et al. 2009; Ghisellini \& Tavecchio 2010; Ghisellini 2012; Sikora et al. 2013) 
and the kiloparsec-scale ones (Georganopoulos et al. 2005; Uchiyama et al. 2005; Mehta et al. 2009). 
They have concluded that the kinetic power in AGN jets is carried mainly by protons, although the 
electron/positron pairs can dominate in the number of particles in the jets. 
The jet composition was derived by the dynamics of the internal shock scenario and/or 
the comparison of the total radiated power with bulk kinetic energy of cold electrons.  
Thus, the concrete answers depend on unknown parameters, especially the actual total 
kinetic power $L_{\rm j}$. 
Note that it is essential to consider the thermal component of AGN jets 
in discussing the jet power (e.g., Kino \& Takahara 2008).

Kino, Kawakatu \& Takahara 2012 (hereafter KKT12) proposed a new approach for testing 
plasma composition of AGN jets by using cocoon dynamics, since invisible plasma (low energy 
electrons/positrons and protons) as well as non-thermal particles plays a role for the expansion 
of cocoons (e.g., Hardcastle \& Worrall 2000; Croston \& Hardcastle 2014). 
In particular, we focused on FRII radio galaxies, to avoid the contamination due 
to jet/lobe entrainment process of surrounding medium via the boundary layer 
of FRI jets (Bicknell 1994; Kaiser 2000; Laing \& Bridle 2002; Laing et al. 2006).  
In contrast to FRIs where entrainment occurs via the boundary layer of FRI jets (Bicknell 1994; Kaiser 2000; 
Laing \& Bridle 2002; Laing et al. 2006),  the relativistic hydrodynamic simulations have shown no significant jet 
entrainment for FRIIs (Sheck et al. 2002; Mizuta et al. 2004; Mizuta et al. 2010). 
As a pilot study, in KKT12 we applied this method to Cygnus A, which is one of well-studied FRIIs 
(e.g., Carilli \& Barthel 1996), and found that plenty of electron/positron pairs are required in Cygnus A. 

Our goal of this paper is to investigate  whether not only Cygnus A jets but also jets of other FRIIs 
include a significant amount of electron/positron pairs. For this purpose, to increase the number of 
samples is evidently crucially important for exploring general properties of FRII jets. 
In order to increase samples, in this paper we constrain the minimum number density of electrons/positrons 
by the non-thermal X-ray emission of the lobes, instead of using the particle supply from hot spots. 
On the other hand, the surrounding ICM densities and pressures are essential to estimate the AGN 
jet power and the source age, and in particular total cocoon pressure. 
Although non-thermal X-rays have been detected for nearly a hundred objects, 
good quality observations of surrounding ICM is available only for several FRIIs. 
Thus, we apply this method to four well-examined FRIIs 
(Cygnus A, 3C219, 3C223, and 3C284), for which the total cocoon pressure has been already evaluated 
by Ito et al. (2008) (hereafter I08). Moreover, the jet power and the cocoon size of Cygnus A are smaller 
than the other three FRIIs. 
Thus, by using these four FRIIs, we can examine whether the plasma composition depends on a jet power 
and a size of cocoon, which could be useful to consider the formation mechanism of FRII jets. 

In $\S 2$, we briefly describe the method, problem setting and dynamical estimate of 
the total pressure in the AGN cocoon, following KKT12. 
In $\S 3$, using this method, we investigate the plasma composition of 
four FRII radio galaxies (Cygnus A, 3C 219, 3C 223 and 3C 284). 
Summary is given in $\S 4$.

\section{Method of constraining plasma composition}
\subsection{Basic idea of the method and problem setting}
On the basis of KKT12, we briefly summarize our method to constrain the plasma 
composition in AGN jets. The method can be divided into three steps as follows; 
(i) The total cocoon pressure $P_{\rm c}^{\rm tot}$ is determined by cocoon dynamics 
following I08 where they obtained $P_{\rm c}^{\rm tot}$ via comparison of the expanding 
cocoon model with the observed cocoon shape. 
Note that the estimation of  $P_{\rm c}^{\rm tot}$ is independent of the plasma composition. 
(ii) The total cocoon pressure is expressed by the equation of state as the sum of partial pressure of each plasma 
component and magnetic field. Importantly, in the equation of state $P_{\rm c}^{\rm tot}$ 
depends on the plasma compositions and particle distribution functions; 
e.g.,  if the positrons are absent in cocoons,  the required proton pressure would be larger than 
the total cocoon pressure, while the pure electrons/positrons jets may result in too many 
thermal electrons/positrons which would conflict with X-ray observations.    
We here take the average energy per each particle (electrons/positrons and protons) in 
the cocoon into account. This strongly depends on the interaction between protons and 
electrons/positrons, and thus several representative cases should be considered. 
(iii) The number density of electrons/positrons can be bounded by maximum and minimum 
values derived by the absence of relativistic thermal bremsstrahlung emission 
from the cocoon and the non-thermal emission from the cocoon, respectively. 
By inserting obtained quantities (the number density of electrons/positrons and the average 
energy per each particle) into the equation of state along with the assumption of  charge neutrality condition, 
we can constrain the number density of electrons/positrons and protons.  
In the following sections, we will explain the individual steps in $\S 2.2$, 2.3 and 2.4.

\subsection{Total cocoon pressure $P_{\rm c}^{\rm tot}$}
The over-pressured cocoon model was initially proposed by Begelman and Cioffi (1989) 
in which the dissipated energy of jet bulk motion is the origin of the total 
pressure of cocoon and a cocoon of FR IIs is expected to be over-pressured 
against ICM pressure ($P_{\rm ICM}$) with a significant sideways expansion. 
Therefore, the assumption of $P_{\rm c}^{\rm tot}=P_{\rm ICM}$ is not valid. 
Kino \& Kawakatu (2005) proposed a new method to estimate the total jet power 
and source age, by using a cocoon model which includes the growth of head 
cross-sectional area of the cocoon and the declining mass density of ICM.  
By comparing the revised cocoon model with the actually observed morphology of 
the cocoons, we have already suggested the method of dynamical constraint on 
$P_{\rm c}^{\rm tot}$ (Kino \& Kawakatu 2005; I08). 
Note that the reliability of the expanding cocoon model was well examined 
in Kawakatu and Kino (2006), was supported by the results of relativistic 
hydrodynamical simulations (Scheck et al. 2002; Perucho \& Marti 2007; 
Mizuta et al. 2010) and was applied to various radio lobes (e.g., Machalski et al. 2010).

Following I08, for the given density of ICM,  $\rho_{\rm ICM}$, the total pressure of cocoon, 
$P_{\rm c}^{\rm tot} $and the head cross-sectional area of the cocoon,  
$A_{\rm h}$ are given as functions of the jet power $L_{\rm j}$ and the source age $t_{\rm age}$. 
Thus, from the condition of overpressure, $P_{\rm c}^{\rm tot} \gg P_{\rm ICM}$ (Kino \& Kawakatu; I08),  
and the comparison with observed $A_{\rm h}$, we can dynamically 
constrain $L_{\rm j}$ and $t_{\rm age}$ (see Table2 in  I08). 
Moreover, the energy equation of cocoons is obtained as 
\begin{eqnarray}\label{eq:pc}
\frac{{\hat \gamma}}{{\hat \gamma}-1}
\frac{P_{\rm c}^{\rm tot}V}{t_{\rm {age}}}= 2 L_{\rm j}. 
\end{eqnarray}  
We here assume $\hat \gamma =4/3$, since the cocoon is expected to be dominated 
by relativistic particles for FRIIs (e.g., Begelman and Cioffi 1989, Kino et al .2007). 
Note that for FRIs some evidence for the non-relativistic thermal emission has been reported 
(e.g., Garrington \& Conway 1991; Seta et al. 2013; O'Sullivan et al. 2013; Stawarz et al. 2013), 
but not for FRIIs 
\footnote{From a theoretical point of views, a mass transfer to extended lobes of FRIIs 
may occur (Kaiser et al. 2000). This would happen through Rayleigh-Taylor instability  
in the interface between the lobe and the ambient ISM and/or ICM. However, 
such an instability starts to grow after the jets have stopped supplying the cocoon 
with the energy (e.g., Reynolds, et al. 2002). This process would be less important 
for supersonic expanding cocoons like FRIIs.}. 

Since $P_{\rm c}^{\rm tot}=L_{\rm j}t_{\rm age}/2V$, 
we can dynamically estimate total pressures $P_{\rm c}^{\rm tot}$ 
by measuring the volume of cocoon $V=2(\pi/3){\cal R}^{2}LS^{3}$, 
where  the linear size of the cocoon along the jet axis $LS$, the aspect ratio 
of the cocoon ${\cal R}\equiv l_{\rm c}/LS<1$, where $l_{\rm c}$ is the 
lateral size of the cocoon. Since the observed quantities (i.e., $LS$ and ${\cal R}$, 
$A_{\rm h}$, $\rho_{\rm ICM}$ and $P_{\rm ICM}$) have some uncertainties, 
the actual $P_{\rm c}^{\rm tot}$ is bounded by maximum and minimum values  
\begin{eqnarray}\label{eq:prange}
P_{\rm min}\leq P_{\rm c}^{\rm tot} \leq P_{\rm max}.
\end{eqnarray}

Thus, we can constrain the total pressure of cocoon $P_{\rm c}^{\rm tot}$, 
which includes the partial pressures of non-radiating particles, 
i.e., thermal electrons/positrons and thermal/non-thermal protons.
The estimation of $P_{\rm c}^{\rm tot}$ has been already performed by I08 for 
four FRII galaxies (Cygnus A, 3C 219, 3C 223 and 3C 284), 
and here we adopt these values in I08.  

\subsection{Equation of State in Cocoons}
We first express the total pressure as the sum of the partial pressures, and then 
the partial pressures are evaluated by setting reasonable particle distribution function 
of non-thermal/thermal population and the average energy per particles. 

Firstly, a total cocoon pressure $P_{\rm c}^{\rm tot}$ is generally expressed as 
\begin{equation}\label{eq:p-def}
P_{\rm c}^{\rm tot} = P_{-}^{\rm T}+P_{+}^{\rm T}+P_{\rm p}^{\rm T}
+P^{\rm NT}_{-} + P^{\rm NT}_{+}+ P^{\rm NT}_{\rm p} 
+P_{\rm B},
\end{equation}
where 
$P^{\rm T}_{-}$,
$P^{\rm T}_{+}$,
$P^{\rm T}_{\rm p}$,
$P^{\rm NT}_{-}$,
$P^{\rm NT}_{+}$,
$P^{\rm NT}_{\rm p}$, and
$P_{\rm B}$ are, the partial pressures of 
thermal (T) electrons,
thermal positrons,
thermal protons,
non-thermal (NT) electrons,
non-thermal positrons,
non-thermal protons, and 
a magnetic pressure respectively. 
In relativistic regime, distinction between thermal and non-thermal 
components is not trivial but they can be characterized by average random Lorentz factor. 
In this paper, we suppose that the thermal component has a Maxwellian-like distribution characterized 
by its temperature, while the non-thermal one refers to particles following a power-law or a broken power law 
distribution characterized by the power-law index and minimum and maximum energies, and by break energy 
for a broken power law.   
Since we consider relativistic plasma, thermal particles correspondingly have relativistic temperature.
Details are described below (see also $\S 2.2$ in KKT12). 
Observationally, ratio of magnetic pressure to that of non-thermal electrons/positrons 
is $P_{\rm B}/P^{\rm NT}_{\pm}=0.01-1$ 
 (e.g., Isobe et al. 2002; Isobe et al. 2005; Kataoka \& Stawarz; Croston et al. 2005; Tashiro et al. 1998, 2009; 
Hardcastle \& Croston 2010; Yaji et al. 2010; Isobe et al. 2011; Stawarz et al. 2013; Isobe \& Koyama 2015). 
The total energy stored in the cocoon exceeds the energy derived from 
the minimum energy condition for the energy of radiating non-thermal electrons and magnetic fields 
by a factor $4-310$  for four FRIIs applied in this paper (see Table 3 in I08). It means that the magnetic 
pressure is a minor contribution for cocoon dynamics. Thus,  we here neglect the magnetic pressure in 
a large scale lobe ($>$ 10kpc) . 

\begin{figure}
\begin{center}
\includegraphics[height=6.5cm,clip]{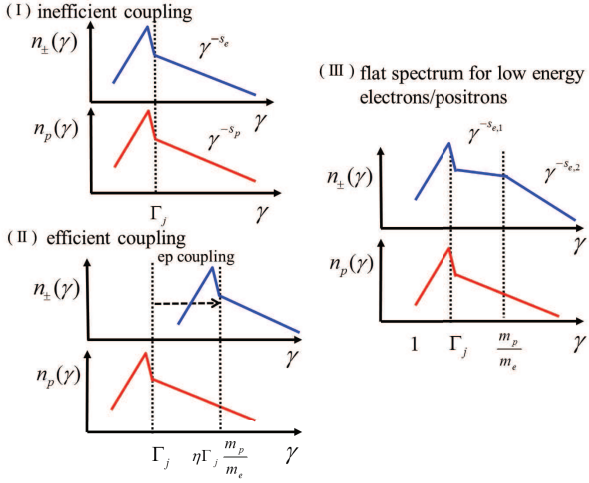}
\end{center}
\caption{
Sketches of the particle distribution function for electrons/positrons and protons 
in cases (I) , (II) and  (III). Case (I) is expected by the standard Fermi acceleration 
which predicts that some fraction of thermal particles are converted into non-thermal one 
with the inefficient coupling between electrons/positrons and protons. 
Case (II)  corresponds to the efficient coupling between electrons/positrons and protons. 
In case (III) non-thermal electrons/positrons obey a broken power law distribution function 
with $s_{\rm e, 1} < 2$ and $s_{\rm e, 2} >2$.
}
\end{figure}

Secondly, we evaluate the partial pressure of each plasma component. 
As a standard case, we suppose a single power-law distribution with a spectral index of 
non-thermal electrons/positrons and protons, $s$, is larger than two, as theoretical work on 
relativistic shocks suggests (e.g., Bednarz and Ostrowski 1998; Kirk et al. 2000; Achterberg et al. 2001; 
Spitkovsky 2008; Sironi and Spitkovsky 2011; Park, Caprioli \& Spitkovsky 2014) and as the radio lobes 
of Cygnus A show $2 < s_{e }< 3$ (e.g., Carilli et al. 1991; Yaji et al. 2010) as well as other FRIIs
 (e.g., Comastri et al. 2003; Croston et al. 2004). 
The particles at the lowest energy are main carrier of non-thermal electron/positron and proton pressures. 
As an additional case, we also consider a broken power-law for non-thermal electrons/positrons. If this is the case, 
the electrons/positrons at the break energy are main carrier of non-thermal electron/positron pressure.  
Using these particle distributions, we can describe the total cocoon pressure as a function of 
five number densities and four average energies per particle using the charge neutrality condition 
(see $\S 2.3.1$). 
Theoretically, the determination of lowest energy for the non-thermal  particles  is one of the most important issues 
because it is directly linked to the injection processes from a thermal pool to a non-thermal component 
(e.g., Spitkovsky 2008; Sironi and Spitkovsky 2011).  
Moreover, the acceleration efficiency of electrons/positrons and protons and the coupling between 
the electrons/positrons and protons are still under debate. Therefore,  we pick up several plausible cases of 
the particle distribution functions and then determine these nine quantities for each case (see cases (I) , (II)  
and (III) in Fig. 1) in $\S 2.3.2$ and $2.3.3$. 

\subsubsection{General properties}
We denote the number densities of protons, $n_{\rm p}$, positrons, $n_{+}$ and electrons, $n_{-}$ as follows: 
\begin{eqnarray} 
n_{\rm p}&=&n_{\rm p}^{\rm T}+n_{\rm p}^{\rm NT}, \nonumber \\
n_{-}&=&n_{-}^{\rm T}+n_{-}^{\rm NT}, \nonumber \\ 
n_{+}&=&n_{+}^{\rm T}+n_{+}^{\rm NT}, \nonumber \\
n_{\pm}&=&n_{-}+n_{+}, 
\end{eqnarray} 
where $n_{\pm}$ is the sum of the total number densities of electrons 
and positrons. Hereafter, the superscripts T and NT represent thermal and 
non-thermal components, respectively. 
In this paper, the number density of electrons, $n_{-}$, positrons, $n_{+}$ and protons, $n_{\rm p}$ 
are related by the charge neutrality conditions; 
\begin{eqnarray} 
n_{+}=n_{-}-n_{\rm p}=(1-\eta)n_{-}, 
\end{eqnarray}
where we define the number density ratio of total protons to total electrons $\eta$ as follows; 
\begin{eqnarray}
\eta & \equiv & \frac{n_{\rm p}}{n_{-}}. 
\end{eqnarray}  
The case of $\eta=0$ corresponds to pure $e^{\pm}$ plasma, while  $\eta=1$ corresponds to the pure $e/p$ plasma. 
Thus, the number of unknown is five. On the average energy (Lorentz factor) per particle for each population, 
we set four quantities as follows, i.e., 
\begin{eqnarray} 
\epsilon_{\pm}^{\rm T}\,\,\,, \epsilon_{\pm}^{\rm NT}\,\,\,, \epsilon_{\rm p}^{\rm T}\,\,\,, \epsilon_{\rm p}^{\rm NT},   
\end{eqnarray} 
where we here suppose $\epsilon_{\pm}^{\rm T}\equiv \epsilon_{+}^{\rm T}=\epsilon_{-}^{\rm T}$ and 
$\epsilon_{\pm}^{\rm NT}\equiv \epsilon_{+}^{\rm NT}=\epsilon_{-}^{\rm NT}$. 
As a result, we have in total nine unknown parameters here. Using these nine quantities, 
we describe the total cocoon pressure, $P_{\rm c}^{\rm tot}$ below. 

For the the standard cases, for the non-thermal populations, we assume the power-law 
distribution functions with $\gamma_{\pm, {\rm min}}\equiv\gamma_{+, {\rm min}}=\gamma_{-,{\rm min}}$,  
$\gamma_{\pm, {\rm max}}\equiv\gamma_{+, {\rm max}}=\gamma_{-,{\rm max}}$, 
$\gamma_{\rm p, {\rm min}}$ and $\gamma_{\rm p, {\rm max}}$. 
\begin{eqnarray}\label{eq:ne}
n_{\pm}^{\rm NT}(\gamma_{\pm}) &\propto& \gamma_{\pm}^{-s_{\rm e}}
(\gamma_{\pm,\rm min}\le \gamma_{\pm} \le \gamma_{\pm,\rm max}), \nonumber \\
n_{\rm p}^{\rm NT}(\gamma_{\rm p}) &\propto& \gamma_{\rm p}^{-s_{\rm p}}
(\gamma_{\rm p,\rm min}\le \gamma_{\rm p} \le \gamma_{\rm p,\rm max}). 
\end{eqnarray}
The spectral indices satisfy $s_{\rm p}=s_{\rm e}\, (>2)$. 
Note that the values of the maximum energy of non-thermal pairs and protons are largely uncertain. 

As an additional case, we also consider a broken power-law distribution function with $\gamma_{\pm, {\rm crit}}
\equiv\gamma_{+, {\rm crit}}=\gamma_{-,{\rm crit}}$. 
\begin{eqnarray}\label{eq:ne}
n_{\pm}^{\rm NT}(\gamma_{\pm}) &\propto& \left \{ 
 \begin{array}{l}
\gamma_{\pm}^{-s_{\rm e, 1}} (\gamma_{\pm,\rm min}\le \gamma_{\pm} \le \gamma_{\pm,\rm crit}), \nonumber \\
\gamma_{\pm, {\rm crit}}^{s_{\rm e, 2}-s_{\rm e, 1}}\gamma_{\pm}^{-s_{\rm e, 2}} 
(\gamma_{\pm,\rm crit}\le \gamma_{\pm} \le \gamma_{\pm,\rm max}), \nonumber \\
 \end{array} \right .
\end{eqnarray}
\begin{eqnarray}\label{eq:ne}
n_{\rm p}^{\rm NT}(\gamma_{\rm p}) &\propto& \gamma_{\rm p}^{-s_{\rm p}}
(\gamma_{\rm p,\rm min}\le \gamma_{\rm p} \le \gamma_{\rm p,\rm max}), 
\end{eqnarray}
where $s_{\rm e, 1} <2 $ and $s_{\rm e, 2}=s_{\rm p} > 2$ are assumed. 
This model may be realized due to the absorption of electron-magnetic waves emitted at the harmonics of 
cyclotron frequency of cold protons (e.g., Hoshino et al. 1991; Amato \& Aronos 2006). 
The average energy of non-thermal electrons, $\epsilon_{\pm}^{\rm NT}$ is determined by 
the critical Lorentz factor $\gamma_{\pm, {\rm crit}}$. 
Actually, this type of spectra has been shown to well explain the observed spectra at hot spots of 
Cygnus A (Stawarz et al. 2007), blazars (e.g., Kang et al. 2014) and Cen A (e.g., Abdo et al. 2010). 

Thus, the total pressure in the cocoon is given by 
\begin{eqnarray}
P_{\rm c}^{\rm tot}&=&P_{\rm c}^{\rm T}+P_{\rm c}^{\rm NT} \nonumber \\
&=&\frac{1}{3}\left[(n_{-}^{\rm T}+n_{+}^{\rm T})\epsilon_{\pm}^{\rm T}m_{\rm e}c^{2}
+n_{p}^{\rm T}\epsilon_{\rm p}^{\rm T}m_{p}c^{2}\right] \nonumber \\
&+&\frac{1}{3}\left[ (n_{-}^{\rm NT}+n_{+}^{\rm NT})\epsilon_{\pm}^{\rm NT}m_{\rm e}c^{2}
+n_{\rm p}^{\rm NT}\epsilon_{\rm p}^{\rm NT}m_{\rm p}c^{2} \right], 
\end{eqnarray}
where the first terms and second terms in the square brackets correspond to the partial pressure of 
electrons/positrons and protons, respectively. Here we assume $n_{\pm}^{\rm NT}=\int_{\gamma_{\pm, \rm min}}^{\gamma_{\pm, \rm max}}n_{\pm}^{\rm NT}(\gamma_{\pm}) d\gamma_{\pm}$ and $n_{\rm p}^{\rm NT}=\int_{\gamma_{\rm p, \rm min}}^{\gamma_{\rm p, \rm max}}
n_{\rm p}^{\rm NT}(\gamma_{\rm p}) d\gamma_{\rm p}$.

On the other hand, by using eq. (1) and the particle (electrons/positrons and protons) number conservation, 
the total cocoon pressure is expressed as 
\begin{eqnarray}
P_{\rm c}^{\rm tot}=\frac{1}{4}\Gamma_{\rm j}m_{\rm e}c^{2}\left(n_{\pm}+\frac{m_{\rm p}}{m_{\rm e}}n_{\rm p}\right)
=\frac{1}{4}\Gamma_{\rm j}m_{\rm e}c^{2}\left[(2-\eta)+\eta\frac{m_{\rm p}}{m_{\rm e}}\right]n_{-}, 
\end{eqnarray}
where $\Gamma_{\rm j}$ is the bulk Lorentz factor.
When we specify particle distribution functions, four average energy per particle ($\epsilon_{\pm}^{\rm T}\,, 
\epsilon_{\pm}^{\rm NT}\,, \epsilon_{\rm p}^{\rm T}\,,\epsilon_{\rm p}^{\rm NT}$) can be determined. 
By using eqs. (10) and (11), we will estimate $\gamma_{\pm, {\rm min}}$, $\gamma_{\pm, {\rm crit}}$ and 
$\gamma_{{\rm p, min}}$, for plausible particle distribution functions in $\S 2.3.2$ and $2.3.3$. 

\subsubsection{Standard distribution function with inefficient $e^{\pm}$/p coupling : case (I)}
We consider the situation that the terminal shock first converts the bulk population of particles 
into thermal ones with Maxwellian-like distribution and accelerates some of them from the thermal pool 
via the first-order Fermi acceleration with the power-law distribution.  
For the first case, which we regard the canonical case, we investigate the equation of state for  the inefficient coupling 
between electrons/positrons and protons (case (I)). 
In this case,  the average energy of thermal electrons/positrons and protons are the same as the minimum 
energy of non-thermal ones, i.e.,  
\begin{eqnarray} 
\epsilon_{\pm}^{\rm T}=\epsilon_{\rm p}^{\rm T}=\epsilon_{\pm}^{\rm NT}=\epsilon_{\rm p}^{\rm NT}. \nonumber
\end{eqnarray} 
Using eqs. (10) and (11), the average energies are determined by the bulk Lorentz factor of AGN jets, $\Gamma_{\rm j}$, 
which is supported by the current particle-in cell simulations (e.g., Spitkovsky 2008; Sironi and Spitkovsky 2011). 
If this is the case, $\epsilon_{\pm}^{\rm T}\,, \epsilon_{\rm p}^{\rm T}\,, \epsilon_{\pm}^{\rm NT}\,, 
\epsilon_{\rm p}^{\rm NT}$ are generally supposed as
\begin{eqnarray} 
\epsilon_{\pm}^{\rm T}&\approx &\frac{3}{4}\Gamma_{\rm j}, \nonumber \\
\epsilon_{\rm p}^{\rm T}&\approx&\frac{3}{4}\Gamma_{\rm j}, \nonumber \\
\epsilon_{\pm}^{\rm NT}&\approx&\gamma_{\pm,\rm min}\approx \frac{3}{4}\Gamma_{\rm j}, \nonumber \\
\epsilon_{\rm p}^{\rm NT}&\approx&\gamma_{\rm p,\rm min}\approx \frac{3}{4}\Gamma_{\rm j}, 
\end{eqnarray}
which are expected when protons and electrons/positrons are separately heated and accelerated 
at termination shocks with a bulk Lorentz factor of jets $\Gamma_{\rm j}$.
Thus, for this case to be self-consistent, the number density and pressure of thermal particles 
should be larger than those of the non-thermal particles. 
However, for general considerations in the followings we here define a free parameter 
$f_{\rm T}\equiv {n_{-}^{\rm T}}/n_{-}^{\rm NT}$ including the case of $f_{\rm T}=0$, 
which denotes the ratio of the number density of thermal electrons to that of non-thermal ones. 
We note that $n_{-}=n_{-}^{\rm NT}$ (i.e., $n_{-}^{\rm T}=0$) corresponds to $f_{\rm T}=0$, 
while  $f_{\rm T}$ is larger than 1 if the number density of thermal electrons is larger than that of 
the non-thermal electrons (see case (I)  in Fig.1).  

Taking into account above assumptions with eq. (10), the total pressure of cocoons, 
$P_{\rm c}^{\rm tot}$ is simply obtained as a function of $\eta$, $f_{\rm T}$ and $n_{-}^{\rm NT}$ as follows. 
\begin{eqnarray}
P_{\rm c}^{\rm tot}(\eta, f_{\rm T}, n_{-}^{\rm NT}) 
=(1+f_{\rm T})\frac{\Gamma_{\rm j}m_{\rm e}c^{2}}{4}\left[(2-\eta)+\eta\frac{m_{\rm p}}{m_{\rm e}}\right]n_{-}^{\rm NT},  \nonumber \\
= 2.7\times 10^{-6}(1+f_{\rm T})n_{-}^{\rm NT}\left[(2-\eta)+\eta\frac{m_{\rm p}}{m_{\rm e}}\right] \left(\frac{\Gamma_{\rm j}}{10}\right)\,
{\rm erg \, cm^{-3}}. 
\end{eqnarray} 
We here define $\eta_{\rm eq}\equiv2/(m_{\rm p}/m_{\rm e}-1)
=1.1\times 10^{-3}$ ($P_{\pm}=P_{\rm p}$). When $\eta > \eta_{\rm eq}$, the proton pressure is dominated 
in the cocoon, while the pair pressure supported cocoon is expected for  $\eta \leq  \eta_{\rm eq}$.
In eq.(13), $\Gamma_{\rm j}$ is a given theoretical parameter, and $n_{-}^{\rm NT}$ and $f_{\rm T}$ (i.e., $n_{-}^{\rm T}$) 
are independently evaluated observationally by assuming the average energy of thermal particles is 
the same as the minimum energy of non-thermal population (see Fig. 1 and $\S 2.4$). 
As a result, we can determine the allowed range of $\eta$ and constrain the plasma composition, comparing the 
equation of state (eq. (13)) with the allowed range of cocoon pressure (eq. (2)). 
As far as we consider the inefficient coupling between electrons/positrons and protons, 
Eq. (13) is valid even for $f_{\rm T} <1$.

As a final remark, we shortly mention the case when the total pressure is dominated by non-thermal particles 
but with the inefficient coupling between electrons/positrons and protons. For this case, too, 
the lowest energy of electrons/positrons and protons are obtained by 
$\gamma_{\pm, {\rm min}}\simeq \gamma_{\rm p\,, min}\simeq \Gamma_{\rm j}$. 
Thus, the formulation of partial pressure of non-thermal components is completely the same as 
the canonical case (eq. (13)) with $f_{\rm T}=0$.  

\subsubsection{Non-canonical cases : cases (II) and (III)}
\begin{itemize}
\item{\it Efficient $e^{\pm}/p$ coupling case: case (II)} \\
For the second case, we here examine the equation of state for  the efficient coupling between electrons/positrons and protons (case (II)). 
In this case, the condition of equipartition is written as  
\begin{eqnarray}
\epsilon_{\pm}^{\rm T}m_{\rm e}c^{2}=\epsilon_{\rm p}^{\rm T}m_{\rm p}c^{2}, 
\,\, \epsilon_{\pm}^{\rm NT}m_{\rm e}c^{2}=\epsilon_{\rm p}^{\rm NT}m_{\rm p}c^{2}.  \nonumber
\end{eqnarray}
We also assume that $\epsilon_{\pm}^{\rm T}=\epsilon_{\pm}^{\rm NT}$ and  $\epsilon_{\rm p}^{\rm T}=\epsilon_{\rm p}^{\rm NT}$ 
(see case (II) for Fig. 1). 
If this is the case, $\epsilon_{\pm}^{\rm T}\,, \epsilon_{\rm p}^{\rm T}\,, \epsilon_{\pm}^{\rm NT}\,, 
\epsilon_{\rm p}^{\rm NT}$ are obtained by using  eqs. (10) and (11); 
\begin{eqnarray}
\epsilon_{\pm}^{\rm T}&\approx &\frac{3\Gamma_{\rm j}}{8}\left[(2-\eta)+\eta \frac{m_{\rm p}}{m_{\rm e}}\right], \nonumber \\
\epsilon_{\rm p}^{\rm T}&\approx &\frac{3\Gamma_{\rm j}}{8}\left[\eta+(2-\eta)\frac{m_{\rm e}}{m_{\rm p}}\right], \nonumber \\
\epsilon_{\pm}^{\rm NT}&\approx&\gamma_{\pm,\rm min} \approx \frac{3\Gamma_{\rm j}}{8}\left[(2-\eta)+\eta \frac{m_{\rm p}}{m_{\rm e}}\right], \nonumber \\
\epsilon_{\rm p}^{\rm NT}&\approx &\gamma_{\rm p,\rm min} \approx \frac{3\Gamma_{\rm j}}{8}\left[\eta+(2-\eta)\frac{m_{\rm e}}{m_{\rm p}}\right].
\end{eqnarray} 
We here note that $\epsilon_{\pm}^{\rm T}=\epsilon_{\pm}^{\rm NT}=(3/4)\Gamma_{\rm j}$ for $\eta=0$, while higher $\epsilon_{\pm}^{\rm T}=
\epsilon_{\pm}^{\rm NT}\gg \Gamma_{\rm j}$ is produced if  $\eta$ is larger because of the efficient electron/proton coupling. 
Taking into account above assumptions with eq. (10), the total cocoon pressure is given by 
\begin{eqnarray}
P_{\rm c}^{\rm tot}(\eta, f_{\rm T}, n_{-}^{\rm NT})
=(1+f_{\rm T})\frac{\Gamma_{\rm j}m_{\rm e}c^{2}}{4}\left[(2-\eta)+\eta\frac{m_{\rm p}}{m_{\rm e}}\right]n_{-}^{\rm NT}. 
\end{eqnarray} 
This formula is the same as case (I), but in this case, $\gamma_{\pm, {\rm min}}$ can be much larger than $\Gamma_{\rm j}$ because of 
$\gamma_{\pm, {\rm min}}\simeq (3/8)\eta\Gamma_{\rm j}{m_{\rm p}/m_{\rm e}}=6.8\times 10^{2}\eta\Gamma_{\rm j}$. 
Thus, $n_{-\,, {\rm min}}\approx n_{-}^{\rm NT}$ is smaller as $\gamma_{\pm,\rm min}$ increases because of $n_{-}^{\rm NT} \propto 
\gamma_{-, {\rm min}}^{1-s_{e}}$ ($s_{e} >2$). 

Comparing this with case (I),  the number density of non-thermal electrons/positrons is smaller. 
Thus, smaller $n_{\rm p}$ may suffice to reproduce the observed cocoon morphologies. 
For a pure electron/proton plasma ($\eta=1$),  the thermal electrons/positrons have higher $\gamma_{\pm, {\rm min}}=
6.8\times 10^{2}\Gamma_{\rm j}$. However,  thermal population dominance ($f_{\rm T} > 1$) can be ruled out because 
an expected big thermal bump is not observed at $\sim 10^{9}$ Hz, i.e., $f_{\rm T}\simeq 0$. Since the difference between 
the two cases is maximal for $\eta=1$, we will especially examine both $\eta=1$ and $f_{\rm T}=0$, 
i.e., higher  $\gamma_{\pm, {\rm min}}=6.8\times 10^{2}\Gamma_{\rm j}$ for case (II). 
Observationally, such a higher $\gamma_{\pm, {\rm min}}\gg \Gamma_{\rm j}$ has been  suggested for the non-thermal electrons 
in several radio galaxies (e.g., Harris et al. 2000; Hardcastle, Birkinshaw, and Worral 2001; Blundell et al. 2006; Godfrey et al. 2009). 
However,  we note that the model spectra with $\gamma_{\pm, \rm min}\geq 2000$ might conflict with 
the observation for Cygnus A (Kino \& Takahara 2004) and that for blazars with $5 < \gamma_{\pm, \rm min} < 160$ 
(Kang et al. 2014) is reported. \\

\item{\it Flat spectrum for low energy electrons/positrons: case (III)}\\
We consider an additional case of  the broken power-law distribution function of non-thermal electrons/positrons (case (III)). 
The difference from case (I) is higher average energy of non-thermal electrons/positrons ($\epsilon_{\pm}^{\rm NT}$) 
and lower  number density of non-thermal electrons/positrons ($n_{\pm}^{\rm NT}$), 
since the electrons at a break energy $\gamma_{\pm, \rm crit}$ dominate (see eq. (9) and Fig. 1). 
According to Stawarz et al. (2007), we assume that $\gamma_{\pm, \rm crit}\simeq m_{\rm p}/m_{\rm e}$. 
If this is the case, $\epsilon_{\pm}^{\rm T}\,, \epsilon_{\rm p}^{\rm T}\,, \epsilon_{\pm}^{\rm NT}\,, 
\epsilon_{\rm p}^{\rm NT}$ are obtained by using eqs. (10) and (11) as follows; 
\begin{eqnarray} 
\epsilon_{\pm}^{\rm T}&\approx &\frac{3}{4}\Gamma_{\rm j}, \nonumber \\
\epsilon_{\rm p}^{\rm T}&\approx&\frac{3}{4}\Gamma_{\rm j}, \nonumber \\
\epsilon_{\pm}^{\rm NT}&\approx&\gamma_{\pm,\rm crit}\approx \frac{m_{\rm p}}{m_{\rm e}}, \nonumber \\
\epsilon_{\rm p}^{\rm NT}&\approx&\gamma_{\rm p,\rm min}\approx \frac{3}{4}\Gamma_{\rm j}. 
\end{eqnarray}
Considering above assumptions with eq. (10), the total cocoon pressure is obtained as  
\begin{eqnarray}
P_{\rm c}^{\rm tot}(\eta, f_{\rm T}, n_{-}^{\rm NT})
=(A_{\pm,\rm b}+f_{\rm T})\frac{\Gamma_{\rm j}m_{\rm e}c^{2}}{4}\left[(2-\eta)+\eta\frac{m_{\rm p}}{m_{\rm e}}\right]n_{-}^{\rm NT},  
\end{eqnarray} 
where $A_{\pm, \rm b}=\gamma_{\pm, \rm crit}/\gamma_{\pm \rm min}=2.4\times 10^2 (\Gamma_{\rm j}/10)$. 
The number density of non-thermal electrons/positrons, $n_{-}^{\rm NT}$, is 
$(\gamma_{\pm, \rm crit}/\gamma_{\pm, \rm min})^{-s_{\rm e, 1}+1}$ 
times smaller than that of case (I). 
Therefore, the total cocoon pressure ($P_{\rm c}^{\rm tot}$) is larger by a factor $(\gamma_{\pm, \rm crit}/
\gamma_{\pm, \rm min})^{-s_{\rm e, 1}+2}$ 
than that of case (I) if the thermal electrons/positrons are negligible, i.e., $f_{\rm T}\ll A_{\pm, \rm b}$.   

\end{itemize}

\subsection{Estimation of electron/positron and proton number density}
First, we estimate the number density of non-thermal electrons/positrons, $n_{\pm}^{\rm NT}$.  
The energy density of non-thermal electrons/positrons and magnetic fields 
can be determined from only observed quantities such as the synchrotron 
radiation flux density, the  inverse Compton X-ray flux density, and the spectral 
index for the radio spectra, $\alpha=(s_{\rm e}-1)/2$. 
We estimate a CMB boosted IC component in accordance 
with Harris \& Grindlay (1979) for 3C219, 3C223 and 3C284, because the 
energy density of CMB photons, $u_{\rm CMB}=4.1\times 10^{-13}(1+z)^{4}\,{\rm erg}\,{\rm cm}^{-3}$ 
is dominant near the radio lobes. 
The energy density of non-thermal electrons/positrons, $u_{\pm}^{\rm NT}$ is given by 
\begin{eqnarray}\label{}
u_{\pm}^{\rm NT}=2.89\times 10^{5}\frac{4\pi d_{\rm L}^{2}(1.1\times 10^{3})^{s_{\rm e}}}
{G(s_{\rm e})(2-s_{\rm e})V(1+z)^{2}}S_{\rm IC} \nonumber \\
\left(\frac{\epsilon_{\rm IC}}{\rm keV}\right)^{\frac{s_{\rm e}}{2}}
\left(\gamma_{\pm, {\rm max}}^{2-s_{\rm e}}-\gamma_{\pm, {\rm min}}^{2-s_{\rm e}}\right), 
\end{eqnarray}
where $d_{\rm L}$, $S_{\rm IC}$ and $\epsilon_{\rm IC}$ are the luminosity distance, the 
IC flux density in Jy at the observed frequency and the energy of the CMB-boosted IC 
X-ray photons, respectively. Here $G(s_{\rm e})\approx 0.5$ in the range of $2 < s_{\rm e} <3$ (see Table 1). 
Note that $s_{\rm e}$ and $\gamma_{\pm, {\rm min}}$ should be replaced by $s_{\rm e, 2}$ and 
$\gamma_{\pm, {\rm crit}}$, respectively for case (III). 
For Cygnus A, we adopt $u_{\pm}^{\rm NT}$ derived by the synchrotron and synchrotron self-Compton (SSC) 
and CMB boosted IC model (Table 7 in Yaji et al. 2010). The observed radio and X-ray information is  summarized in Table 1. 

In general, the number density and energy density of non-thermal electrons/positrons are obtained as 
\begin{eqnarray}\label{}
n_{\pm}^{\rm NT}&=&\int_{\gamma_{\pm, \rm min}}^{\gamma_{\pm, \rm max}}n_{\pm}^{\rm NT}(\gamma_{\pm}) d\gamma_{\pm}, \nonumber \\
u_{\pm}^{\rm NT}&=&m_{\rm e}c^{2}\int_{\gamma_{\pm, \rm min}}^{\gamma_{\pm, \rm max}}
n_{\pm}^{\rm NT}(\gamma_{\pm}) d\gamma_{\pm} \nonumber.
\end{eqnarray}
For cases (I) and (II),  the number density of non-thermal electrons/positrons can be expressed by 
\begin{eqnarray}\label{eq:nNT}
n_{\pm}^{\rm NT}=\frac{s_{e}-2}{s_{e}-1}
\frac{u_{\pm}^{\rm NT}}{m_{e}c^{2}}\gamma_{\pm,{\rm min}}^{-1}. 
\end{eqnarray}
On the other hand, $n_{\pm}^{\rm NT}$ for case (III) is given by 
\begin{eqnarray}\label{eq:nNT}
n_{\pm}^{\rm NT}=\frac{(s_{e, 2}-2)(s_{e,1}-2)}{(s_{e,1}-1)(s_{e,1}-s_{e,2})}
\frac{u_{\pm}^{\rm NT}}{m_{e}c^{2}}\gamma_{\pm,{\rm min}}^{-1}
\left(\frac{\gamma_{\pm, {\rm crit}}}{\gamma_{\pm, {\rm min}}} \right)^{s_{e, 1}-2}. 
\end{eqnarray}

The minimum energy of non-thermal electrons/positrons is treated as mentioned in 
$\S 2.3$. 
On the other hand, we adopt the maximum energy of  non-thermal pairs as $\gamma_{\pm,\rm max}=10^{5}$. 
Although the values of $\gamma_{\pm,\rm max}$ are largely uncertain, it is reasonable 
to suppose that $\gamma_{\pm,\rm max}\gg \gamma_{\pm,\rm min}$. 
Note that $n_{\pm}^{\rm NT}$ is sensitive to the uncertainty of $s_{\rm e}$ and $S_{\rm IC}$, 
but the main results ($\S 3$) does not change significantly even if we take into account of these 
uncertainties. 
 
Next, we constrain the upper limit of the number density of thermal electrons/positrons, $n_{\pm}^{\rm T}$ 
by the absence of thermal bremsstrahlung from the relativistic electrons/positrons in the cocoons/
lobes viewed in $X$-ray observations (Wilson et al. 2000, 2006). 
This estimation can be applied only for case (I). For case (II), 
we do not use the upper limit of  $n_{\pm}^{\rm T}$ because of $n_{\pm}\simeq n_{\pm}^{\rm NT}$. 
The observed  $X$-ray emissions associated with radio lobes are 
non-thermal emissions and there is no evidence for the relativistic thermal $X$-ray
emission from cocoons/lobes for FRIIs (Harris and Krawczynski 2006 for review), though non-relativistic 
thermal X-ray emission was detected for a few FRI radio galaxies (e.g., Seta et al. 2013; 
O'Sullivan et al. 2013; Stawarz et al. 2013). 
From this, we can safely use the condition of 
$L_{{\rm X, obs}}>L_{\rm brem}(n_{\pm}^{\rm T},T_{\pm})$ 
where 
$L_{\rm brem}/V=
\alpha_{\rm f}r_{\rm e}^{2}m_{\rm e}c^{3}(n_{\pm}^{\rm T})^{2}
F_{\pm}(\Theta_{\pm})~{\rm erg~s^{-1}~cm^{-3}}$, 
$F_{\pm}(\Theta_{\pm})=
48\Theta_{\pm}(\ln 1.1\Theta_{\pm}+5/4)$, and 
$\Theta_{\pm}=kT_{\pm}/m_{e}c^{2}\simeq \gamma_{\pm, {\rm min}}$,
for bremsstrahlung at relativistic temperature (Eq. (22) in Svensson 1982) and $\alpha_{\rm f}$ and $r_{e}$ are the 
fine structure constant and the classical electron radius, 
respectively. From this, we obtain the maximum  $n_{\pm, {\rm max}}^{\rm T}$ as follows:
\begin{eqnarray}\label{eq:Lx}
n_{\pm, {\rm max}}^{\rm T} =
\left(\frac{L_{\rm X, obs}}
{V\alpha_{\rm f}r_{e}^{2}m_{e}c^{3}
F_{\pm}(\Theta_{\pm})}\right)^{1/2}.
\end{eqnarray}
Once $n_{\pm}=n_{+}+n_{-}$ is estimated, the number density of electrons/positrons 
can be determined as 
\begin{eqnarray}
&& n_{-,{\rm min}} \leq n_{-} \leq n_{-, {\rm max}}, \nonumber \\
&& (1-\eta^{\rm NT})n_{-,{\rm min}} \leq n_{+} \leq (1-\eta^{\rm T})n_{-, {\rm max}} \nonumber
\end{eqnarray}
where $n_{-,{\rm min}}=n_{\pm}^{\rm NT}/(2-\eta^{\rm NT})$, 
$n_{-,{\rm max}}=n_{\pm, {\rm max}}^{\rm T}/(2-\eta^{\rm T})$, 
$\eta^{\rm NT}=n_{\rm p}^{\rm NT}/n_{-}^{\rm NT}$ and 
$\eta^{\rm T}=n_{\rm p}^{\rm T}/n_{-}^{\rm T}$. 
The values of $\eta^{\rm T}$ and $\eta^{\rm NT}$ do not affects on the 
estimate of $n_{-, {\rm min}}$ and $n_{-, {\rm max}}$ significantly because of $0\le  \eta^{\rm T}  \le 1$ 
and $0\le  \eta^{\rm NT}  \le 1$.
Thus, we adopt $\eta=\eta^{\rm T}=\eta^{\rm NT}$ for simplicity.
Note that $\eta=\eta^{\rm T}=\eta^{\rm NT}$ when the acceleration efficiency 
of electrons/positrons is the same as that of protons.
Finally, the number density of protons is obtained as 
\begin{eqnarray}\label{eq:nproton}
&& n_{\rm p}=\eta n_{-},  \nonumber
\end{eqnarray}
As the final step, the allowed ranges of $n_{\rm p}$ (or $\eta$) can be estimated  
by adjoining the range of $P_{\rm c}^{\rm tot}$ (see eqs. (13), (15) and(17)). 

\section{Results}
We apply the above method to four FRII radio galaxies 
(Cygnus$\,$A, 3C$\,$219, 3C$\,$223, and 3C$\,$284), 
for which the total cocoon pressures are available 
in the previous work (I08). 
As for the bulk Lorentz factor of jets, $\Gamma_{\rm j}$,  
in concordance to the superluminal motion measured by VLBI (Very Long Baseline Interferometer), 
we suppose that the jet is relativistic at $> 10\,{\rm kpc}$ with $3\leq \Gamma_{\rm j} \leq 30$ 
(e.g., Krichbaum et al. 1998; Lister et al. 2001; Bach et al. 2002; Kellerman et al. 2004). 
On the other hand, we adopt the observable quantities, i.e.,the cocoon morphology $\cal{R}$, 
the cocoon volume $V$, and the total cocoon pressure $P_{\rm c}^{\rm tot}$ for individual objects. 
Using these quantities, we can constrain the allowed region of $n_{-}, n_{+}$ and $n_{\rm p}$. 
In $\S 3.1-3.4$, we show the allowed regions of $n_{-}$ and $n_{\rm p}$ for the inefficient $e^{\pm}/p$ 
coupling case (case (I)) for each FRII (Cygnus$\,$A, 3C$\,$219, 3C$\,$223, 3C$\,$284). 
In $\S 3.5$, we examine the efficient $e^{\pm}/p$ coupling case (case (II)) and 
the case of a broken power-law distribution function of non-thermal electrons/positrons (case(III)) 
for all four FRII radio galaxies.  

\subsection{Application to Cygnus$\,$A: case (I)}
In our previous work (KKT12), we have already investigated the plasma composition for Cygnus A 
by using basically the same method as the present work. In KKT12, the minimum number density of 
electrons/positrons is constrained by using the particle supply from hot spots. Instead of this, 
we here adopt much simpler method using radio lobe emissions, because this method is easily 
applied to other FRII radio galaxies with the observed spectra of radio lobes, and thus we can increase 
the number of sample. To compare with other objects, 
it is necessary to re-examine the plasma composition of jets in Cygnus A by using the present 
method described in $\S 2$. 

\begin{figure}
\begin{center}
\includegraphics[height=6.5cm,clip]{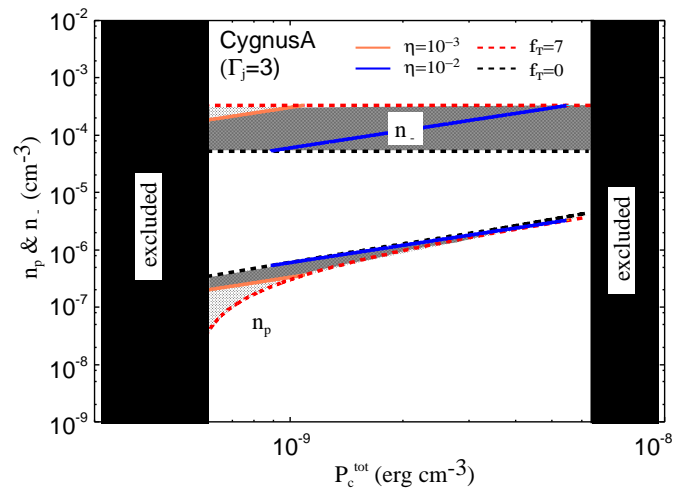}
\end{center}
\caption{
Allowed range of $n_{-}$ (upper region) and $n_{\rm p}$ (lower region) for 
Cygnus A with $\Gamma=3$ (the high-$n$ case).
The allowed range of $P_{\rm c}^{\rm tot}$ is $6\times 10^{-10}\,{\rm erg}\,{\rm cm}^{-3}\leq 
P \leq 6.5\times 10^{-9}\,{\rm erg}\,{\rm cm}^{-3}$. 
The several colored lines represent the different values of $\eta$, 
e.g., $\eta=10^{-3}$ (orange) and $\eta=10^{-2}$ (blue). 
The region in which $P_{\pm} > P_{\rm p}$ holds is colored in light gray, 
while the region where $P_{\pm} < P_{\rm p}$ is satisfied is colored in dark gray.
We find that the number density of electrons/positrons is always larger than $n_{\rm p}$.  
}
\end{figure}

\begin{figure}
\begin{center}
\includegraphics[height=6.5cm,clip]{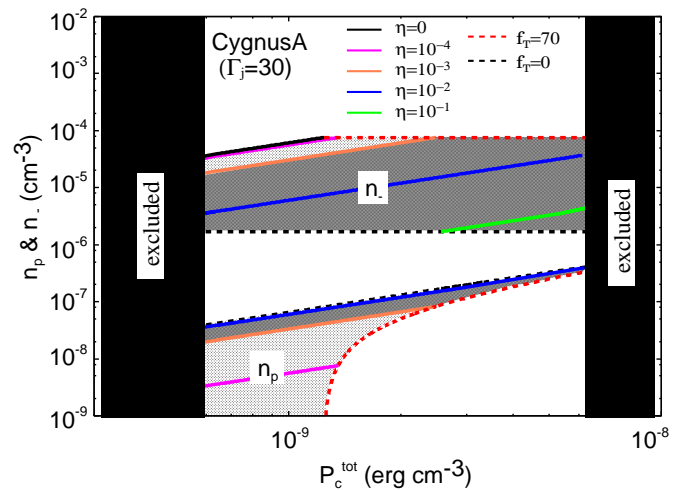}
\end{center}
\caption{
Same as Fig. 2 but with $\Gamma=30$ (the low-$n$ case.). 
Though the obtained region of $n_{-}$ 
and $n_{\rm p}$ are about 30 times smaller than the one in Fig.1 
because of $n_{-,{\rm min}}\propto \Gamma^{-1.4}$, the basic results 
are same as Fig. 2. 
}
\end{figure}

On the cocoon morphology, we can evaluate the range of cocoon aspect ratio, 
$0.25 \leq {\cal R} \leq 0.5$ from the radio images of Cygnus A (Carilli et al. 1991; Wilson et al. 2002, 
2006; Lazio et al. 2006; Yaji et al. 2010; Chon et al. 2013). 
Using $V=1\times 10^{70}{\cal R}^{2}\,{\rm cm}^{3}$, we derived 
the total pressure  $P_{\rm c}^{\rm tot}$ as 
\begin{eqnarray}
6.0\times 10^{-10}\,{\rm erg}\,{\rm cm}^{-3} \leq P_{\rm c}^{\rm tot} \leq 6.4\times 10^{-9}\,{\rm erg}\,{\rm cm}^{-3} \nonumber. 
\end{eqnarray}
The range of $P_{\rm c}^{\rm tot}$ is determined by the uncertainty of $\cal{R}$, 
$L_{\rm j}$ and $t_{\rm age}$. 
From eq. (19),  the number density of non-thermal electrons/positrons  is obtained as 
$n_{\pm}^{\rm NT}=2.75\times 10^{-4}\gamma_{\pm, \rm min}^{-1.4}\,{\rm cm}^{-3}$. 
We note that the pressure of non-thermal electrons/positrons is obtained as 
\begin{eqnarray}
P_{\pm}^{\rm NT}=1.0\times 10^{-9}\gamma_{\pm\, ,\rm min}^{-0.4}{\cal R}_{0.25}^{2}\,{{\rm erg}\, {\rm cm}^{-3}} \nonumber 
\end{eqnarray}
with ${\cal R}_{0.25}={\cal R}/0.25$, which is one order of magnitude larger than the ambient pressure at the head of lobes, 
$P_{\rm a}=8\times 10^{-11}\,{{\rm erg}\, {\rm cm}^{-3}}$ (Arnaud et al. 1984). 
For $\Gamma_{\rm j}=3$, the non-thermal pressure  is $P_{\pm}^{\rm NT}=6.5\times 10^{-10}
\,{{\rm erg}\, {\rm cm}^{-3}}$, which is comparable to $P_{\rm min}$, while 
the non-thermal pressure $P_{\pm}^{\rm NT}=2.5\times 10^{-10}\,{{\rm erg}\, {\rm cm}^{-3}}$, is 
two times smaller than $P_{\rm min}$ for $\Gamma_{\rm j}=30$. Thus, the additional pressures 
(thermal electrons/positrons and/or thermal/non-thermal protons) would be needed. 
By using the upper limit of thermal bremsstrahlung emission at X-ray band, 
i.e., $2\times 10^{-13}\,{\rm erg}\,{\rm s}^{-1}\,{\rm cm}^{-2}$ (Yaji et al. 2010), 
we estimate the maximal number density of thermal electrons/positrons, 
$n_{\pm, {\rm max}}^{\rm T}=7.81\times 10^{-4}\gamma_{\pm, \rm min}^{-0.5}\,{\rm cm}^{-3}$ (eq. (21)). 
Considering the uncertainties of bulk Lorentz factor $\Gamma_{\rm j}$, we here 
investigate two limiting cases with $\Gamma_{\rm j}=3$ and $\Gamma_{\rm j}=30$.  
Because of $n_{-}^{\rm NT}\propto \gamma_{\pm, \rm min}^{-1.4}\propto \Gamma_{\rm j}^{-1.4}$, 
the number density of electrons with $\Gamma_{\rm j}=3$ is about 30 times larger than that 
with $\Gamma_{\rm j}=30$. Thus, we call the case of  $\Gamma_{\rm j}=3$ 
the high-$n$ case, and that of $\Gamma_{\rm j}=30$ the low-$n$ case, respectively. 

Figure 2 shows the region of $n_{-}$(upper region in Fig. 2) and $n_{\rm p}$ (lower region in Fig. 2) 
for the high-$n$ case, i.e., lower Lorentz factor $\Gamma_{\rm j}=3$. 
As we discussed in  $\S2.4$, $n_{-}$ is bounded by $n_{-, \rm min}$ and $n_{-, \rm max}$. 
Once, $n_{-}$ is constrained, the number density of positrons and that of protons are given by 
$n_{+}=(1-\eta)n_{-}$ and $n_{\rm p}=\eta n_{-}$, respectively.  
For given $\eta$, the minimum/maximum pressure in the cocoon can be also described by 
the fraction of thermal electrons $f_{\rm T}$. Using the parameter $f_{\rm T}$, we will discuss 
how the fraction of invisible thermal components affects on the matter contents of AGN jets.
The upper boundary of the electron number density, $n_{-}=n_{-, {\rm max}}^{\rm T}$,  
corresponds to $f_{\rm T}=7$ (red dashed line).  
Corresponding to this, the lower boundary of $n_{\rm p}$ is determined with the red dashed lines. 
On the other hand, the lower limit of  $n_{-}=n_{-}^{\rm NT}$ is $f_{\rm T}=0$ (black dashed line). 
Corresponding upper boundary of $n_{\rm p}$ is shown by the black dashed line.  
Finally, the allowed regions of $n_{-}$ and $n_{\rm p}$ can be obtained by adjoining the range of 
$P_{\rm c}^{\rm tot}$ (see eq. (2)). Thus, the allowed regions are the shaded regions in Fig. 2.  
As seen in Fig. 2, the allowed region of $n_{-}$ and $n_{\rm p}$ can be divided into several regions 
with different $\eta=n_{\rm p}/n_{-}$. Several colored solid lines correspond to the cocoon 
pressure with a different $\eta$. 

In Fig. 2, we find that the number density of electrons/positrons ($n_{\pm}$) is much larger 
than that of protons, since $\eta \leq$ $7 \times 10^{-2}$ for the permitted range of cocoon pressure. 
This means that plenty of positron mixture is required, in other words, the pure electron/proton plasma 
is firmly ruled out.  We further examine the partial pressures of electrons/positrons and protons. 
The regions with $\eta < \eta_{\rm eq}$ (light-gray regions) show the one in which $P_{\pm} > P_{\rm p}$ 
holds, while the region of $\eta \geq \eta_{\rm eq}$ (dark-gray regions) represents the one in which $P_{\pm} 
\leq P_{\rm p}$ holds. As seen in Fig.2, both the pair and electron/proton $(e/p)$ plasma pressure supported 
cocoons are allowed in the range of $P_{\rm c}$. 
On the contribution of thermal electrons, the larger $f_{\rm T}$ results in  the lower $\eta$ 
for fixed $P_{\rm c}^{\rm tot}$. For $f_{\rm T} =7$, both $e^{\pm}$ pair plasma pressure 
supported and $e/p$ pressure supported cocoons are allowed (i.e., $10^{-4} < \eta < 10^{-2}$), 
while the $e/p$ plasma pressure supported cocoons ($\eta > \eta_{\rm eq}$) are obtained 
in the absence of thermal electrons ($f_{\rm T}=0$). If $f_{\rm T}=0$, the range of $\eta$ is 
$7\times 10^{-3} \leq \eta \leq 7\times 10^{-2}$.

Figure 3 displays the result for the low-$n$ case (i.e., higher Lorentz factor $\Gamma_{\rm j}=30$). 
Similar to the high-$n$ case, $n_{\pm} > n_{\rm p}$ always holds with $\eta \leq$ $0.4$ 
for the permitted range of cocoon pressure. This means that the pure  electron/proton jets are 
ruled out. The number density of electrons is about 30 times smaller than that for the 
high-$n$ case because of the decrease in $n_{-}^{\rm NT}$ due to larger $\gamma_{-,\,\rm min}$. 
Corresponding to this, the upper boundary of $n_{\rm p}$ becomes larger, and thus the maximal value of 
$\eta$ is larger than that for the high -$n$ case. 
In addition, it is found that both ${\rm e/p}$ plasma and $e^{\pm}$ pair pressure supported cocoon are 
allowed within the current observations. 
The dependence on $f_{\rm T}$ is same as the high-n case, but the pure $e^{\pm}$ pair plasma is 
also allowed for $f_{\rm T}\geq 30$. 
It is also important to mention that the CMB boosted IC optical/infrared emission may be detected 
because plenty of the thermal electrons with $\gamma_{\pm,{\rm min}}=30$ exist in this case.  
If we take the maximum $n_{\pm, {\rm max}}^{\rm T}=10^{-4}\,{\rm cm}^{-3}$, 
the flux density at $1.5\times 10^{14}\,{\rm Hz}$ is obtained as 
$F_{\rm IC}\simeq 10^{-13}\,{\rm erg}\,{\rm s}^{-1}\,{\rm cm}^{-2}$. 
\footnote
{
The energy loss rate by IC scattering of CMB photons is 
$P_{\rm IC}=\frac{4}{3}\sigma_{\rm T}c\gamma_{\pm, \rm min}^{2}u_{\rm CMB}$ where $\sigma_{\rm T}$ is the Thomson cross-section.
The IC frequency is $\nu_{\rm IC}=\gamma_{\pm, {\rm min}}^{2}\nu_{\rm CMB}$ with $\nu_{\rm CMB}=1.6\times 10^{11}(1+z)\,{\rm Hz}$. 
The total luminosity of IC component in lobes is described as 
$L_{\rm IC}\simeq VP_{\rm IC}n_{\pm}^{\rm T}$. 
Then, the observed flux density 
at $\nu_{\rm IC}=1.5\times 10^{14}\,{\rm Hz}$ is obtained as 
$F_{\rm IC}=\frac{L_{\rm IC}}{4\pi d_{\rm L}^{2}}=10^{-9}n_{\pm}^{\rm T}\,{\rm erg}\,{\rm s}^{-1}\,{\rm cm}^{-2}$. 
}
This value is comparable to the upper limit of flux density for the Cygnus A hotspots observed by previous optical/infrared 
observations (Meisenheimer et al. 1997). Thus, it is interesting to explore whether the CMB boosted IC optical/infrared emission 
from cocoon is prominent or not by future observations.

Finally, we shortly comment on the results of Cygnus A in KKT12. Comparing the present results (Figs. 2 and 3) 
with the previous work (Figs. 4 and 5 in KKT12) , we find that the main results coincide with each other. 

\subsection{Application to 3C 219: case (I)}
Unlike Cygnus A other objects lack low frequency radio image (74 MHz and 0.3 GHz) , 
so that it is generally hard to measure the aspect ratio of cocoons ${\cal R} $ from 
radio images, since the emission of cocoons which are far from the hot spots 
is very faint at GHz frequency because of synchrotron cooling. 
Therefore, we here explore a wide range of $0.5 \leq {\cal R} \leq 1$, taking into account 
the large uncertainty on the shape of cocoons. 
Using $V=5.2\times 10^{71}{\cal R}^{2}\,{\rm cm}^{3}$, we derive the total pressure of 
cocoon, $P_{\rm c}^{\rm tot}$ as 
\begin{eqnarray}
6.3\times 10^{-11}\,{\rm erg}\,{\rm cm}^{-3} \leq P_{\rm c}^{\rm tot} \leq 1.2\times 10^{-9}\,{\rm erg}\,{\rm cm}^{-3} \nonumber. 
\end{eqnarray}
This range is determined by the uncertainty of $\cal{R}$, $L_{\rm j}$ 
and $t_{\rm age}$.  
From eq. (19), the number density of non-thermal electrons/positrons is given by 
$n_{\pm}^{\rm NT}=3.1\times 10^{-5}\gamma_{\pm, \rm min}^{-1.6}\,{\rm cm}^{-3}$. 
The pressure of non-thermal electrons is obtained as 
\begin{eqnarray}
P_{\pm}^{\rm NT}=2\times 10^{-11}\gamma_{\pm, \rm min}^{-0.6}{\cal R}^{-2}\,{{\rm erg}\, {\rm cm}^{-3}} \nonumber,
\end{eqnarray}
which is one order of magnitude larger than the ambient pressure at the head of lobes, 
$P_{\rm a}=1.6\times 10^{-12}\,{{\rm erg}\, {\rm cm}^{-3}}$ (I08). 
Since the non-thermal $e^{\pm}$ pair pressure is at least 1.5 times smaller than the minimum of $P_{\rm c}^{\rm tot}$,  
i.e., $P_{\pm}^{\rm NT}=4.1\times 10^{-11}\,{{\rm erg}\, {\rm cm}^{-3}}$ for $\Gamma_{\rm j}=3$ and  
$P_{\pm}^{\rm NT}=1.0\times 10^{-11}\,{{\rm erg}\, {\rm cm}^{-3}}$ for $\Gamma_{\rm j}=30$, 
the additional pressure is required. 
By using the upper limit of thermal bremsstrahlung emission at X-ray band (Comastri et al. 2003; Croston 2004), 
we estimate the upper limit of number density of thermal electrons/positrons, 
$n_{\pm, {\rm max}}^{T}=1.99\times 10^{-4}\gamma_{\pm, {\rm min}}^{-0.5}\,{\rm cm}^{-3}$ (eq. (21)). 
In the same way as Cygnus A, we consider two limiting cases with $\Gamma_{\rm j}=3$ and 
$\Gamma_{\rm j}=30$.  
Because of $n_{\pm}^{\rm NT}\propto \gamma_{\pm, \rm min}^{-1.6}\propto \Gamma_{\rm j}^{-1.6}$, 
the number density of electrons with $\Gamma_{\rm j}=3$ is about 30 times larger than that 
with $\Gamma_{\rm j}=30$. 

Figure 4 shows the allowed region of $n_{-}$ and $n_{\rm p}$ for the high-$n$ case 
with lower Lorentz factor $\Gamma_{\rm j}=3$. As mentioned in $\S 2.3$, the total 
pressure of cocoon is a function of $\eta$, and then the colored lines correspond 
to the cocoon pressure with different $\eta$. 
It is found that $\eta \leq$ $10^{-1}$  is required to reproduce permitted range of 
cocoon pressure. This suggests that positron mixture is required, 
in other words, the pure electron/proton plasma is firmly ruled out.  
For partial pressure,  both the pair and ${\rm ep}$ plasma 
pressure supported cocoons are allowed in the range of $P_{\rm c}^{\rm tot}$.  
For $f_{\rm T}$=0, the range of $\eta$ is $7\times 10^{-3} \leq \eta \leq 10^{-1}$, 
so that the pair pressure supported cocoon would be ruled out. 
On the other hand, for $f_{\rm T} =14$, the range of $\eta$ is $0\leq \eta \leq 10^{-2}$, 
i.e., the electron/positron pressure supported cocoon is also permitted.

\begin{figure}
\begin{center}
\includegraphics[height=6.5cm,clip]{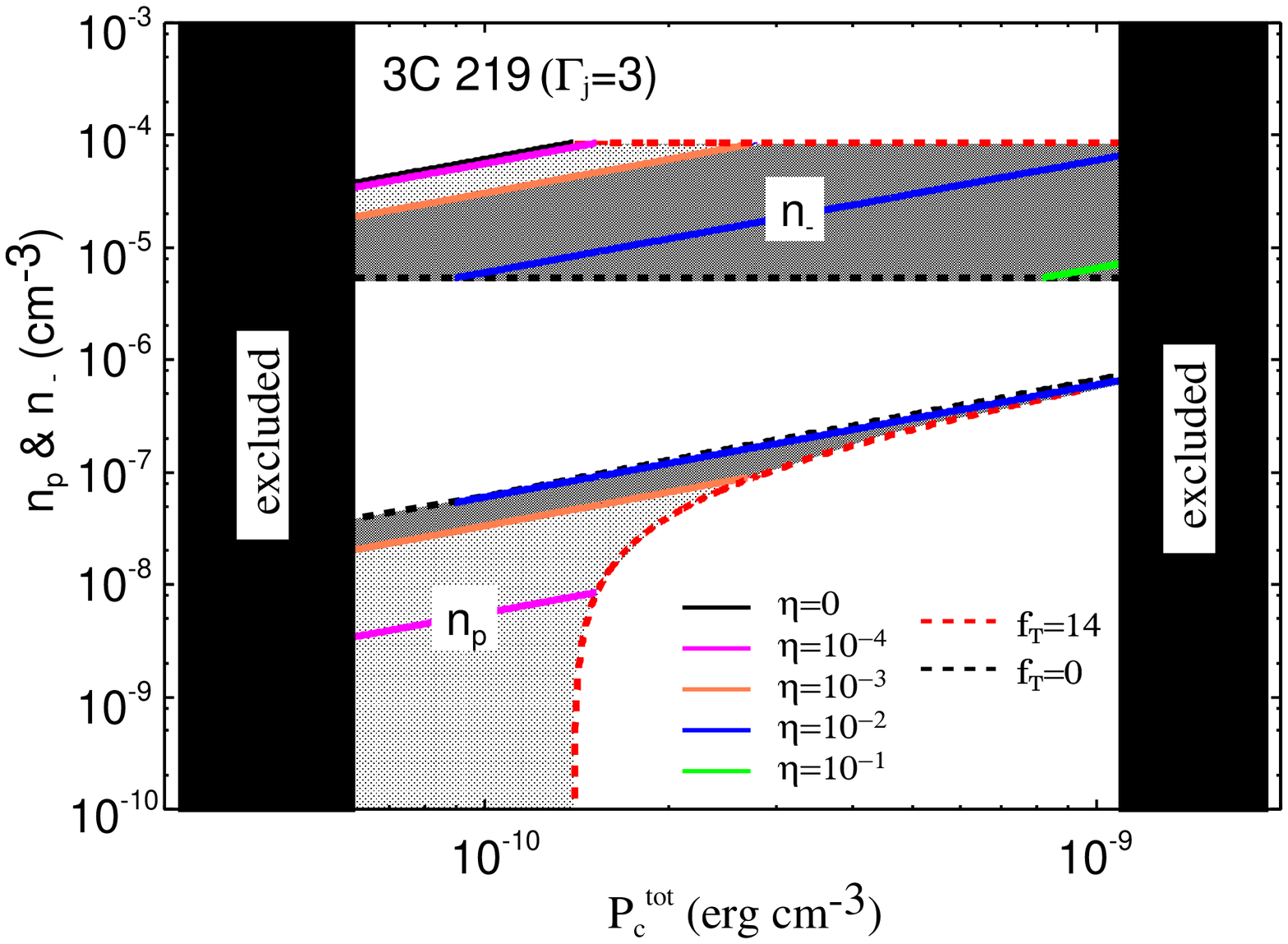}
\end{center}
\caption{
Same as Fig. 2 but for 3C 219.
}
\end{figure}
\begin{figure}
\begin{center}
\includegraphics[height=6.5cm,clip]{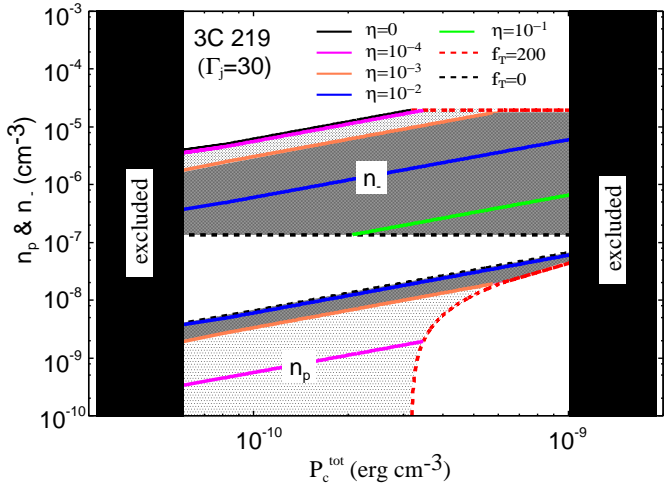}
\end{center}
\caption{
Same as Fig. 3 but for 3C 219.
}
\end{figure}

Figure 5 displays the result for the low-$n$ case (i.e., higher Lorentz factor $\Gamma_{\rm j}=30$). 
Similar to the high-$n$ case, $n_{\pm} > n_{\rm p}$ always holds and $\eta \leq$ $0.4$ is required 
for the permitted range of cocoon pressure. This suggests that the pure electron/proton jets are 
ruled out.The number density of electrons is two orders of magnitude smaller than that for the high-$n$ case.
In addition, it is found that both $e/p$ plasma and $e^{\pm}$ pair pressure supported cocoons are 
allowed within the current observations. 
In this case,  If the partial pressure of thermal electrons (e.g., $f_{\rm T}>30$) is dominant, 
the lepton number density dominated jets with $\eta \leq 5\times 10^{-3}$ are favorable.  
From Figs. 4 and 5, we find that the overall results are quite similar to Cygnus A,  
even if the jet kinetic power of 3C219 is one order of magnitude larger than that 
of Cygnus A (Table 2 in I08). 

\subsection{Application to 3C 223: case (I)}
In the same way as 3C219, we here explore a wide range of $0.5 \leq {\cal R} \leq 1$, 
taking into account of the large uncertainty on the shape of cocoons (Orru et al. 2010) . 
Using $V=2\times 10^{72}{\cal R}^{2}\,{\rm cm}^{3}$, we derive the total pressure 
$P_{\rm c}^{\rm tot}$ as 
\begin{eqnarray}
1.5\times 10^{-12}\,{\rm erg}\,{\rm cm}^{-3} \leq P_{\rm c}^{\rm tot} 
\leq 6.5\times 10^{-11}\,{\rm erg}\,{\rm cm}^{-3} \nonumber.
\end{eqnarray}

This  range is determined by the uncertainty of $\cal{R}$, $L_{\rm j}$ and $t_{\rm age}$. 
From eq. (19), the number density of non-thermal electrons/positrons is given by 
$n_{\pm}^{\rm NT}=1.0\times 10^{-6}\gamma_{\pm, {\rm min}}^{-1.5}\,{\rm cm}^{-3}$. 
The pressure of non-thermal electrons/positrons is obtained as 
\begin{eqnarray}
P_{\pm}^{\rm NT}=8.3\times 10^{-13}\gamma_{\pm, {\rm min}}^{-0.5}{\cal R}^{-2}\,{{\rm erg}\, {\rm cm}^{-3}} \nonumber,
\end{eqnarray}
which is comparable to the ambient pressure  $P_{\rm a}=1.2\times 10^{-12}\,{{\rm erg}\, {\rm cm}^{-3}}$ (I08). 
For $\Gamma_{\rm j}=3$, the non-thermal pressure  is $P_{\pm}^{\rm NT}=1.9\times 10^{-12}
\,{{\rm erg}\, {\rm cm}^{-3}}$, which is comparable to $P_{\rm min}$, while 
the non-thermal pressure $P_{\pm}^{\rm NT}=6.3\times 10^{-13}\,{{\rm erg}\, {\rm cm}^{-3}}$, is 
three times smaller than $P_{\rm min}$ for $\Gamma_{\rm j}=30$. 
Thus, the additional pressures is necessary except for lowest $P_{\rm c}^{\rm tot}$. 
By using the upper limit of thermal bremsstrahlung emission at X-ray band (Croston 2004), 
we estimate the upper limit of number density of thermal electrons/positrons 
$n_{\pm, {\rm max}}^{\rm T}=3.0\times 10^{-5}\gamma_{\pm, {\rm min}}^{-0.5}\,{\rm cm}^{-3}$ (eq. (21)). 
We here consider two limiting cases with $\Gamma_{\rm j}=3$ and $\Gamma_{\rm j}=30$.  
Because of $n_{\pm}^{\rm NT}\propto \gamma_{\pm, \rm min}^{-1.5}\propto \Gamma_{\rm j}^{-1.5}$, 
the number density of electrons with $\Gamma_{\rm j}=3$ is about 30 times larger than that 
with $\Gamma_{\rm j}=30$. 

Figure 6 shows the allowed region of $n_{-}$ and $n_{\rm p}$ for the high-$n$ case, 
i.e., lower Lorentz factor $\Gamma_{\rm j}=3$. As mentioned in $\S 2.3$, the total 
pressure of cocoon is a function of $\eta$, and then the colored lines correspond 
to the cocoon pressure with different $\eta$. 
We find that $\eta \leq$ $0.2$ is required to provide the permitted range of cocoon pressure. 
This suggests that the pure electron/proton plasma is firmly ruled out.  
For partial pressure,  as seen for Cygnus A and 3C219, both the pair and $e/p$ plasma 
pressure supported cocoons are permitted in the range of $P_{\rm c}^{\rm tot}$. 
If non-thermal pressure ($f_{\rm T}=0$) is dominant, the range of $\eta$ is 
$5\times 10^{-2} \leq \eta \leq 0.2$, so that the pair pressure supported 
cocoon is rejected. On the other hand, for $f_{\rm T} >3$, both the pair 
and $e/p$ plasma pressure supported cocoons are allowed. 

\begin{figure}
\begin{center}
\includegraphics[height=6.5cm,clip]{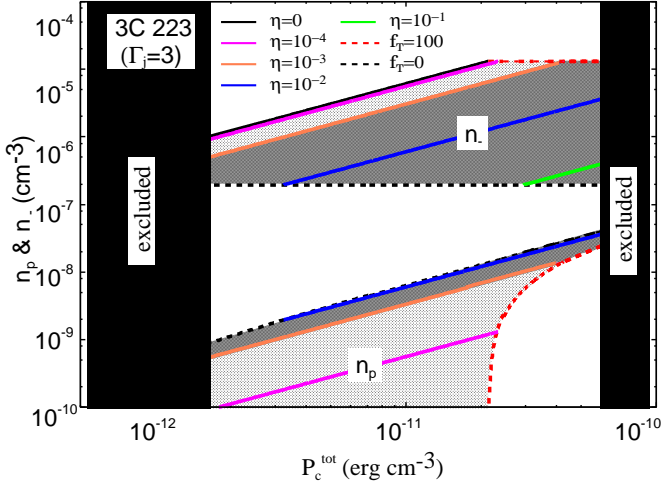}
\end{center}
\caption{
Same as Fig. 2 but for 3C 223.
}
\end{figure}

\begin{figure}
\begin{center}
\includegraphics[height=6.5cm,clip]{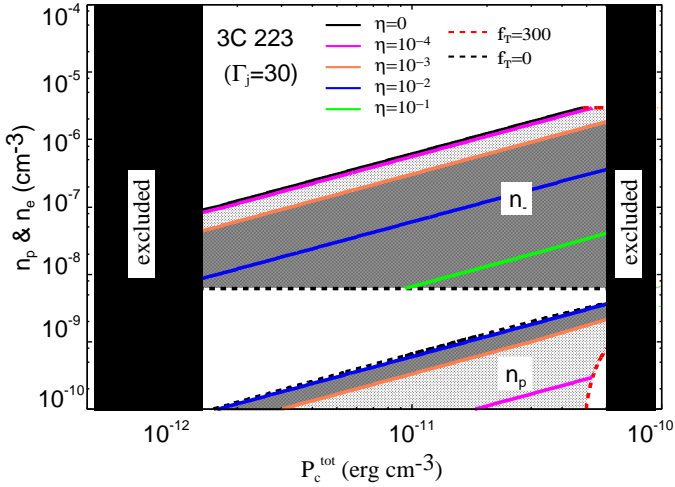}
\end{center}
\caption{
Same as Fig. 3 but for 3C 223.
}
\end{figure}

Figure 7 displays the result for the low-$n$ case (i.e., higher Lorentz factor $\Gamma_{\rm j}=30$). 
Similar to the high-$n$ case, $n_{\pm} > n_{\rm p}$ always holds and $\eta \leq 0.5$  is required to provide 
for permitted range of cocoon pressure. This suggests that pure electron/proton jets is ruled out. 
The number density of electrons is about 30 times  smaller than that for the high-$n$ case. 
Additionally, we find that both $e/p$ plasma and $e^{\pm}$ pair pressure supported cocoons are 
allowed within the current observations. If the partial pressure of thermal electrons/positrons is dominant 
(e.g., $f_{\rm T} >10 $), the electrons/positrons pressure supported cocoon is favorable.  
From Figs. 6 and 7, we find that the overall results are quite similar to Cygnus A, 
in other words, the mixture of positrons is needed in the relativistic jets of 3C223. 

\subsection{Application to 3C 284: case (I)}
As is for the other two FRIIs (3C219 and 3C223), we here explore a wide range of $0.5 \leq {\cal R} \leq 1$, 
taking into account of the large uncertainty on the shape of cocoons. 
Using $V=5.5\times 10^{71}{\cal R}^{2}\,{\rm cm}^{3}$,  we estimate the total pressure  
$P_{\rm c}$ as 
\begin{eqnarray}
9.5\times 10^{-12}\,{\rm erg}\,{\rm cm}^{-3} \leq P_{\rm c}^{\rm tot} \leq 3.8\times 10^{-9}\,{\rm erg}\,{\rm cm}^{-3} \nonumber. 
\end{eqnarray}
From eq. (19), the number density of non-thermal electrons/positrons is given by 
$n_{\pm}^{\rm NT}=6.4\times 10^{-5}\gamma_{\pm, {\rm min}}^{-1.9}\,{\rm cm}^{-3}$. 
The pressure of non-thermal electrons/positrons is obtained as 
\begin{eqnarray}
P_{\pm}^{\rm NT}=4\times 10^{-11}\gamma_{\pm, {\rm min}}^{-0.9}{\cal R}^{-2}\,{{\rm erg}\, {\rm cm}^{-3}} \nonumber,
\end{eqnarray} 
which is two orders of magnitude larger than the ambient pressure  $P_{\rm a}=
(3.7-6.4)\times 10^{-13}\,{{\rm erg}\, {\rm cm}^{-3}}$ (I08). 
Since the non-thermal $e^{\pm}$ pair pressure is smaller than the minimum of $P_{\rm c}^{\rm tot}$,  
i.e., $P_{\pm}^{\rm NT}=6.0\times 10^{-11}\,{{\rm erg}\, {\rm cm}^{-3}}$ for $\Gamma_{\rm j}=3$ and  
$P_{\pm}^{\rm NT}=7.5\times 10^{-12}\,{{\rm erg}\, {\rm cm}^{-3}}$ for $\Gamma_{\rm j}=30$, 
the additional pressure is required. 
By using the upper limit of thermal bremsstrahlung emission at X-ray band (Croston 2004), 
we estimate the maximal number density of thermal electrons/positrons 
$n_{\pm, {\rm max}}^{\rm T}=1.9\times 10^{-4}\gamma_{\rm \pm, {\rm min}}^{-0.5}\, {\rm cm}^{-3}$ (eq. (21)). 
As well as other FRIIs, we consider two limiting cases with $\Gamma_{\rm j}=3$ and $\Gamma_{\rm j}=30$.  
Because of $n_{-}^{\rm NT}\propto \gamma_{-, \rm min}^{-1.9}\propto \Gamma_{\rm j}^{-1.9}$, 
the number density of electrons with $\Gamma_{\rm j}=3$ is about two orders of magnitude 
larger than that that with $\Gamma_{\rm j}=30$. 

Figure 8 shows the allowed region of $n_{-}$ and $n_{\rm p}$ for the high-$n$ case 
with lower Lorentz factor $\Gamma_{\rm j}=3$. As mentioned in $\S 2.3$, the total 
pressure of cocoon is a function of $\eta$, and then the colored lines correspond 
to the cocoon pressure with different $\eta$. 
We find that $\eta \leq$  $3\times 10^{-2}$ is required to provide the permitted range of cocoon pressure. 
This shows that the pure $e/p$ plasma is firmly ruled out again.  
As for the partial pressure,  as seen for Cygnus A, 3C219 and 3C223, both the pair and $e/p$ plasma 
pressure supported cocoons are permitted in the range of $P_{\rm c}^{\rm tot}$.
 If non-thermal pressure ($f_{\rm T}=0$) is dominant, the range of $\eta$ is $0 \leq \eta \leq 3\times 10^{-2}$, 
so that both pair and $e/p$ plasma pressure supported cocoons can be permitted.

Figure 9 displays the result for the low-$n$ case (i.e., higher Lorentz factor $\Gamma_{\rm j}=30$). 
Similar to the high-$n$ case, $n_{\pm} > n_{\rm p}$ always holds and $\eta \leq 0.2$ is required to provide 
the permitted range of cocoon pressure. This shows again that pure $e/p$ jets can be 
rejected. The number density of electrons is two orders of magnitude smaller than that for the high-$n$ case 
because of the decrease of $n_{-}^{\rm NT}$. 
Moreover, it is found that both $e/p$ plasma and $e^{\pm}$ pair pressure supported cocoons are 
allowed within the current observations. 
If non-thermal pressure ($f_{\rm T}=0$) is dominant, the range of $\eta$ is 
$5\times 10^{-3} \leq \eta \leq 0.4$, so that the $e/p$ plasma pressure supported jets 
is accepted. On the other hand, if  plenty of thermal electrons/positrons exist (e.g.,  $f_{\rm T} >20$), 
both pair and $e/p$ plasma pressure supported cocoons are permitted. 

From Figs. 8 and 9, we find that the overall results are quite similar to Cygnus A, 
in other words, the mixture of positrons is needed in the relativistic jets of 3C284.

\subsection{Application to four FRIIs: cases (II) and (III)}
Let us consider electron/proton pressure supported cocoons with 
high $\gamma_{\pm, {\rm min}}\gg \Gamma_{\rm j}$ (case (II)). 
Here we especially examine whether the case of pure non-thermal electron/proton plasma ($\eta=1$ and $f_{\rm T}=0$) 
is allowed or not for four FRIIs, since this possibility is not included in the canonical case, as shown in $\S 3.1-3.4$. 
Substituting the relation $n_{-}^{\rm NT}\propto \gamma_{-, \, {\rm min}}^{1-s_{\rm e}}$ constrained by radio observations to eq. (15), 
we obtain the total cocoon pressure $P_{\rm c}^{\rm tot}(\eta=1) $ for case (II). 
Here we use spectral indexes at radio frequencies (178-750 MHz) for individual objects (see Table 1) and 
\begin{eqnarray}
\gamma_{\pm, {\rm min}}=(3/8)\Gamma_{\rm j}{m_{\rm p}/m_{\rm e}}=6.8\times 10^{2}\Gamma_{\rm j}. \nonumber
\end{eqnarray}
As a result, for all four FRIIs radio galaxies (Cygnus A, 3C 219, 3C 223 and 3C 284), 
we find that the total cocoon pressure for this case $P_{\rm c}^{\rm tot}(\eta=1) $ is 
well below $P_{\rm min}$ which is derived by cocoon dynamics (see Table 2). 
Because of $P_{\rm p}^{\rm NT}\propto n_{-}^{\rm NT}$,  $P_{\rm c}^{\rm tot}(\eta=1) $ cannot reach the actual 
total cocoon pressure (see eq. (2)). Therefore,  we conclude that the pure $e/p$ plasma 
composition ($\eta=1$) is ruled out even for case (II). 

\begin{figure}
\begin{center}
\includegraphics[height=6.5cm,clip]{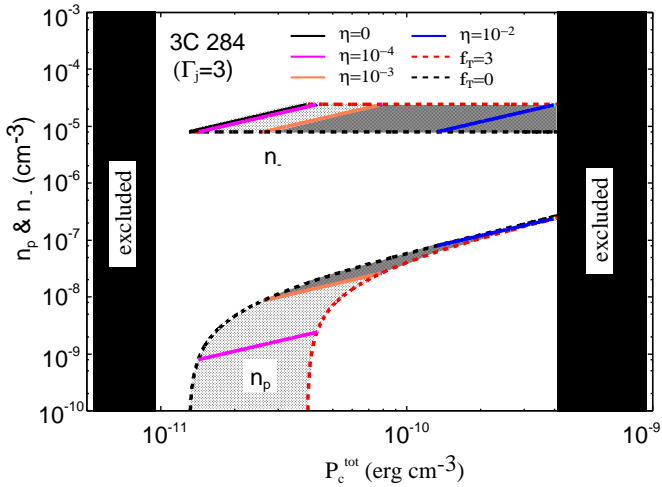}
\end{center}
\caption{
Same as Fig. 2 but for 3C 284.
}
\end{figure}

\begin{figure}
\begin{center}
\includegraphics[height=6.5cm,clip]{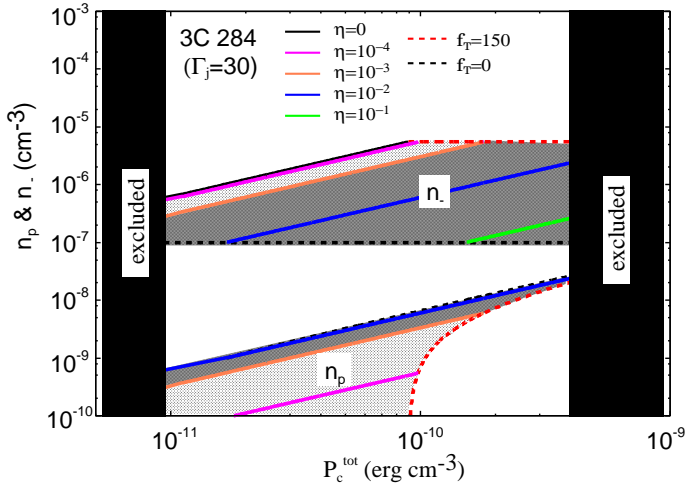}
\end{center}
\caption{
Same as Fig. 3 but for 3C 284.
}
\end{figure}

Lastly, we examine the case of  flat spectrum for low energy non-thermal electrons/positrons (case (III)), 
which has been suggested by some modeling of the non-thermal broad-band emission of lobes and hot spots 
in several radio galaxies (e.g., Stawarz et al. 2007; Abdo et al. 2010; Kang et al. 2014). 
Assuming $s_{\rm e, 1}=1.5$ (Stawarz et al. 2014 for Cygnus A), the number density of  non-thermal electrons/positrons 
($n_{-}^{\rm NT}$) is $0.06 (\Gamma_{\rm j}/10)^{0.5}$ times lower than that  of case (I), 
whereas the average energy of them ($\epsilon_{\pm}^{\rm NT}\simeq \gamma_{\pm, {\rm crit}}$) is 
$2.4\times 10^{2} (\Gamma_{\rm j}/10)$ times larger than case (I). 
Totally, for the same value of $n_{-}^{\rm NT}$, a difference between case (I) and case (III) is the larger total 
cocoon pressure $P_{\rm c}^{\rm tot}$ by a factor $15(\Gamma_{\rm j}/10)^{0.5}$ (see eqs. (13) and (17) in $\S 2.3.3$). 
If we consider the case of pure non-thermal electron/proton plasma ($\eta=1$ and $f_{\rm T}=0$), 
the total cocoon pressure is given by $P_{\rm c}^{\rm tot}(\eta=1)=4.0\times 10^{-2}n_{-}^{\rm NT}(\Gamma_{\rm j}/10)$ 
erg cm$^{-3}$. Considering the allowed range of the number density of non-thermal electrons/positrons,  
the total cocoon pressure is well above $P_{\rm max}$ which is derived by cocoon  dynamics (Table 2). 
This conclusion makes stronger if the thermal electrons/positrons exist ($f_{\rm T} >0 $).
Therefore, though allowed lower value of $\eta$ is slightly larger  than case (I), we find the positron-free plasma 
comprising of protons and electrons is  ruled out for case (III). 

\section{Summary}
In this paper, we have investigated plasma composition of relativistic jets in four FRII radio galaxies  
based on the allowed ranges of total pressures in the cocoons. In KKT12, we found that electron/positron pairs 
always dominate in terms of number density for Cygnus A. 
The main purpose of the present work is to examine whether the significant mixture of electron-positron pair
plasma seen in Cygnus A is also realized in other FR II jets with very different jet powers and source sizes. 
We carefully treat partial pressure of both thermal and non-thermal particles in cocoon, taking account of 
plausible particle distribution functions (i.e., inefficient $e^{\rm}/p$ coupling case, efficient $e^{\rm}/p$ coupling case 
and flat spectrum for low energy electrons/positrons). 
In order to increase the number of FRIIs, we derive the minimum number density of electrons/positrons directly 
from the observed non-thermal X-ray emission of the radio lobes, instead of conducting analysis of hot spot 
emissions as KKT12. Moreover, the surrounding ICM densities and pressures are essential to evaluate 
the total cocoon pressure. Since good quality observations of surrounding ICM are available only for several objects,  
we apply this method to Cygnus$\,$A, 3C$\,$219, 3C$\,$223 and 3C$\,$284, for which the total cocoon pressures 
are available in I08. By inserting obtained maximum and minimum number density of electrons/positrons into 
the equation of state along the assumption of charge neutrality condition, we can constrain the number density of 
protons. Our main results are as follows: 

\begin{itemize}
\item 
We find that the positron-free plasma consisting of protons and electrons can be rejected, 
even if we consider several plausible particle distribution functions. 
In other words, mixtures of electrons/positrons and protons in AGN jets are realized 
for all four FRII radio galaxies, i.e., $n_{\pm}/n_{\rm p} > 2$ {when we consider plausible particle distribution}. 
Thus, it indicates that plenty of electrons/positrons still survive in $>10$ kpc-scale jets. 
Comparing individual properties of four FRIIs, it is also found that the plasma composition 
does not depend on the jet power (its range is $L_{\rm j}\approx 10^{45}\sim10^{47}\,{\rm erg}\,{\rm s}^{-1}$) 
and size of  cocoons (its range is $60\sim 340$ kpc). This may indicate that a significant mixture of 
electrons/positrons is a general feature in FRII jets.

Since these results are closely related to the formation mechanism of relativistic jets, we here compare with 
previous works on the plasma composition and discuss the pair production mechanism. 
For the previous studies of bulk-Compton emission of blazar (e..g, Sikora \& Madejski 2000; Moderski et al. 2004; 
Celotti et al. 2007; Kataoka et al. 2008; Ghisellini \& Tavecchio 2010), the upper limit of $n_{\rm \pm}/n_{\rm p}$ 
is $10-20$, although the uncertainty of $L_{\rm j}$ is large and the scale of jets is different. 
Such comparisons may be interesting to reveal the dependence of plasma compositions on the scale of  AGN jets. 
For the mechanism of pair production, Iwamoto \& Takahara (2002, 2004) and Asano \& Takahara (2007, 2009) 
showed that a sufficient amount of electron positron pairs survive as a relativistic outflow if the temperature of 
Wien fireball at the photosphere is relativistic, where all the cross-sections of pair creation, 
annihilation, and Compton scattering are of the same order. Therefore, a balance between pair creation 
and annihilation is realized, and the number densities of pairs and photons are of the same order 
as long as photons and pairs are coupled with each other.  According to this model, MeV-peak emission 
due to electron/positron pairs annihilation is predicted. Considering the interaction between the accelerated 
protons and MeV-peak photons, ”orphan” TeV flares in blazars may be a possible manifestation of 
mixture of $e^{\pm}$ plasma (e..g, Fraija 2015).  

Theoretically, there has been an argument that electron positron pairs in AGN jets annihilate in the course 
of jet propagations (e.g., Celotti \& Fabian1993; Blandford \& Levinson 1995). 
If copious amount of  electron/positron pairs are confined in a compact region, 
the pair annihilation inevitably occur. Therefore, the electron/positron loading issue should 
be considered more seriously; the work on the pair plasma outflow (Iwamoto \& 
Takahara 2002, 2004; Asano \& Takahara 2007, 2009), on the electrons/positrons pair loading problem 
(e.g.,Blandford \& Levinson 1995; Yamasaki, Takahara \& Kusunose 1999; Sikora \& Madejski  2000) 
and on two-flow paradigm in which jets are composed of electron/positron pair core surrounded by 
a heavier electron/proton sheath (e.g., Sol et al. 1989, Henri et al. 1993; Ghisellini 2012) may be an 
 important starting point to understand the formation of relativistic AGN jets.  

\item  
We find that for  abundant thermal pairs $f_{\rm T} >10$ the $e/p$ plasma pressure supported cocoons 
are  also allowed, while for no thermal electrons ($f_{\rm T}=0$) the electron/proton plasma pressure 
supported cocoons are permitted. As $f_{\rm T}$ increases, the invisible thermal electron pressure 
becomes large and thus the contribution of proton pressure becomes small.  
On the other hand, the protons may be dynamically dominated for FRI radio galaxies as suggested by 
the energetics of their radio lobes (e.g., De Young 2006). 
It might imply that FRI jets are intrinsically different from FRII jets and/or the contamination of protons due to 
entrainment is significant for FRI jets.  
Then, it is worth to explore the emission from the relativistic thermal electrons/positrons in FRII cocoons, 
taking properly into account the electron/positron temperature and the distribution of 
electron/positron number density, such as inverse Compton of CMB photons at IR/optical bands, 
which is left in future work. 
It is also important to examine the luminosity of  a secondary emission produced 
by high energy non-thermal protons (e.g., Atoyan \& Dermer 2008).  Since its luminosity is 
closely linked to $P_{\rm p}^{\rm NT}$, we may constrain the fraction of $P_{\rm p}^{\rm NT}$ 
on $P_{\rm c}^{\rm tot}$. 
 
\end{itemize}

\section*{Acknowledgments} 
We thank the anonymous referee for many valuable comments and suggestions to improve our original manuscript. 
NK acknowledges the financial support of Grant-in-Aid for Young Scientists (B:25800099).

\bsp
\label{lastpage}

\newpage

\begin{table}
\begin{center}
Table 1  Observed radio and X-ray information \\[3mm]
\begin{tabular}{ccccccccc}
\hline \hline
Source & $z$  & $d_{\rm L}$  & LS & $V$  & $s_{\rm e}$  & $S_{\rm IC}(\rm 1keV)$ & $L_{\rm j}$  &references \\
           &       & (${\rm Mpc}$)  & $({\rm kpc})$ & (${\rm cm^{3}}$) &  &(Jy)  &  ($10^{46}{\rm erg}\, {\rm s}^{-1}$) \\
\hline 
Cygnus A &  0.057 & 249  &  60 & $1.0\times 10^{70} {\cal R}^{2}$ & 2.4  & $(5.2\sim7.8)\times 10^{-8}$ &$0.35-2.6$ &1, 4   \\
3C 219    &  0.174   & 829  &  210 & $5.2\times 10^{71}{\cal R}^{2}$ &  2.6  & $(7.0\sim9.0)\times 10^{-9}$&$2.6-43$& 2, 3, 4  \\
3C 223    &  0.137   & 635  & 340 & $2.0\times 10^{72}{\cal R}^{2}$  &  2.5  & $3.1\times 10^{-9}$ &$0.07-2.9$&3, 4  \\
3C 284    &  0.239   & 1182 & 260 & $5.5\times 10^{71}{\cal R}^{2}$ &  2.9  & $(0.9\sim1.9)\times 10^{-9}$ &$0.1-18$&3,4  \\ 
\hline
\end{tabular}
\end{center}
{
Notes.--- Col(1): Names of radio sources. Cols. (2) and (3): Redshift and luminosity
distance, respectively, calculated for the cosmology with $H_{0}$=71 km s$^{-1}$,  
$\Omega_{\rm M}=0.3$, and $\Omega_{\Lambda}=0.7$. 
Cols. (4) and (5): Linear size and volume of cocoons. 
Col. (6): Spectral indexes at radio frequencies (178-750 MHz), which are taken from the Table 1 of Hardcastle et al. (1998). 
Col. (7): Flux density of radio lobes at 1 keV. 
Col. (8): Jet power.  
Col. (9): References \\
references--- (1) Yaji et al. 2010, (2) Comastri et al. 2003, (3) Croston et al. 2004, (4) Ito et al. 2008}
\end{table}

\begin{table}
\begin{center}
Table 2 Results of cases (II) and (III)  \\[3mm]
\begin{tabular}{ccccc}
\hline \hline
Source & Case(II): $P_{\rm c}^{\rm tot}(\eta=1)$ & Case(III): $P_{\rm c}^{\rm tot}(\eta=1)$ & $P_{\rm min}$  & $P_{\rm max}$\\
           & ($\rm erg\, cm^{-3}$)  & ($\rm erg\, cm^{-3}$) & ($\rm erg\, cm^{-3}$)  & ($\rm erg\, cm^{-3}$)\\
\hline 
Cygnus A &  $1.5\times 10^{-11}$ & $4.4\times 10^{-7}$  & $6.0\times 10^{-10}$ & $6.4\times 10^{-9}$ \\
3C 219    &  $7.9\times 10^{-14}$ & $3.1\times 10^{-8}$  & $6.3\times 10^{-11}$ & $1.2\times 10^{-9}$  \\
3C 223    &  $6.8\times 10^{-15}$ & $1.3\times 10^{-9}$  & $1.5\times 10^{-12}$ & $6.5\times 10^{-11}$ \\
3C 284    &  $6.9\times 10^{-13}$ & $8.1\times 10^{-9}$  & $9.5\times 10^{-12}$ & $3.8\times 10^{-9}$ \\ 
\hline
\end{tabular}
\end{center}
{
Note that we calculate the pure electron/proton dominated cocoon pressure $P_{\rm c}^{\rm tot}(\eta=1)$ 
for cases (II) and (III) by assuming $\Gamma_{\rm j}=10$ and $f_{\rm T}=0$. 
}
\end{table}


\begin{thebibliography}{}
\bibitem[Abdo et al. 2010]{Abdo10}
Abdo A.~A., et al., 2010, Sci, 328, 725 
\bibitem[Achterberg et al.(2001)]{Achterberg01} 
Achterberg, A.,Gallant, Y.~A., Kirk, J.~G., \& Guthmann, A.~W.\ 2001, 
MNRAS, 328, 393
\bibitem[Amato\& Arons(2006)]{Amato06} 
Amato, E., \& Arons, J.\ 2006, ApJ, 653, 325
\bibitem[Asano \& Takahara(2007)]{Asano07}
Asano, K., \& Takahara, F.,\ 2007, ApJ, 655, 762
\bibitem[Asano \& Takahara(2009)]{Asano09}
Asano, K., \& Takahara, F.,\ 2009, ApJ, 690, L81
\bibitem[Atoyan \& Dermer(2008)]{Atoyan08}
Atoyan, A., \& Dermer, C.~D.,\ 2008, ApJ, 687, L75
\bibitem[Bach et al.(2002)]{Bach02}
Bach, U., Krichbaum, T.~P., Alef, W., Witzel, A.,
\& Zensus, J.~A.,\ 2002,
Proceedings of the 6th European VLBI Network Symposium,
eds.: E. Ros, R. W. Porcas, A. P. Lobanov,
and J. A. Zensus, p. 155
\bibitem[Bednarz\& Ostrowski(1998)]{Bednarz98} Bednarz, J., \& Ostrowski, M.\ 1998, 
Physical Review Letters, 80, 3911
\bibitem[Begelman et al.(1984)]{Begelman84}
Begelman, M.~C., Blandford, R.~D., \& Rees, M.~J.\ 1984,
Rev.~Mod.~Phys., 56, 255
\bibitem[]{}
Begelman M.~C., Sikora M., 1987, ApJ, 322, 650
\bibitem[Begelman \& Cioffi(1989)]{Begelman89}
Begelman, M.~C., \& Cioffi, D.~F.,\ 1989, ApJL, 345, L21
\bibitem[]{} 
Bicknell G.~V., 1994, ApJ, 422, 542
\bibitem[Blandford \& Znajek(1977)]{Blandford77}
Blandford, R.~D., \& Znajek, R.~L.\ 1977, MNRAS, 179, 433
\bibitem[]{}
Blandford R.~D., Payne D.~G., 1982, MNRAS, 199, 883
\bibitem[Blandford \& Levinson(1995)]{Blandford95}
Blandford, R.~D., \& Levinson, A.,\ 1995,
ApJ, 441, 79
\bibitem[Blundell et al.(2006)]{Blundell06}
Blundell, K.~M., Fabian, A.~C.,
Crawford, C.~S., Erlund, M.~C., \& Celotti, A.,\ 2006, ApJ, 644, L13
\bibitem[Carilli et al.(1991)]{Carilli91}
Carilli, C.~L., Perley, R.~A., Dreher, J.~W., \& Leahy, J.~P.\
1991, ApJ, 383, 554
\bibitem[Carilli \& Barthel(1996)]{Carilli96}
Carilli, C.~L., \& Barthel, P.~D.,\ 1996, A\&ARv, 7, 1
\bibitem[]{1993MNRAS.264..228C} Celotti A., Fabian A.~C., 
1993, MNRAS, 264, 228
\bibitem[]{}
Celotti A., Ghisellini G., Chiaberge M., 2001, MNRAS, 321, L1
\bibitem[]{}
Chon G., B{\"o}hringer H., Krause M., Tr{\"u}mper J., 2012, A\&A, 545, L3 
\bibitem[]{}
Comastri A., Brunetti G., Dallacasa D., Bondi M., Pedani M., Setti G., 2003, MNRAS, 340, L52
\bibitem[]{2007MNRAS.375..417C}
Celotti A., Ghisellini G., Fabian A.~C., 2007, MNRAS, 375, 417
\bibitem[]{}
Croston J.~H., Birkinshaw M., Hardcastle M.~J., Worrall D.~M., 2004, MNRAS, 353, 879
\bibitem[]{}
Croston J.~H., Hardcastle M.~J., Harris D.~E., Belsole E., Birkinshaw M., Worrall D.~M., 2005, ApJ, 626, 733
\bibitem[]{}
Croston J.~H., Hardcastle M.~J., 2014, MNRAS, 438, 3310
\bibitem[De Young (2006)]{DeYoung06}
De Young, D.~S.,\ 2006, ApJ, 648, 200
\bibitem[]{}
Falle S.~A.~E.~G., 1991, MNRAS, 250, 581 
\bibitem[Fabian et al.(2002)]{Fabian02}
Fabian, A.~C., Celotti, A., Blundell, K.~M.,
Kassim, N.~E., \& Perley, R.~A.,\ 2002,
MNRAS, 331, 369
\bibitem[]{}
Fraija N., 2015, arXiv, arXiv:1501.04165 
\bibitem[Garrington\& Conway(1991)]{Garrington91}
Garrington, S.~T., \& Conway, R.~G.\ 1991, MNRAS, 250, 198
\bibitem[Georganopoulos et al.(2005)]{Georganopoulos05}
Georganopoulos, M., Kazanas, D., Perlman, E.,
\& Stecker, F.~W.,\ 2005, ApJ, 625, 656
\bibitem[Ghisellini et al. 2014]{Ghisellini14}
Ghisellini, G. Tavecchio, T., Maraschi, L, Celotti, A., Sbarrato, T. 2014, Nature, 515, 376 (arXiv: 1411.5368) 
\bibitem[Ghisellini \& Tavecchio(2010)]{Ghisellini10}
Ghisellini, G., \& Tavecchio, F.\ 2010, MNRAS, 409, L79
\bibitem[Godfrey et al.(2009)]{Godfrey09}
Godfrey, L.~E.~H., et al.,\ 2009, ApJ, 695, 707
\bibitem[]{}
Godfrey L.~E.~H., Shabala S.~S., 2013, ApJ, 767, 12
\bibitem[]{}
Hardcastle M.~J., Alexander P., Pooley 
G.~G., Riley J.~M., 1998, MNRAS, 296, 445 
\bibitem[]{Hardcastle00} 
Hardcastle, M.~J., \& Worrall, D.~M.,\ 2000,MNRAS, 319, 562
\bibitem[Hardcastle et al.(2001)]{Hardcastle01}
Hardcastle, M.~J., Birkinshaw, M., \& Worrall, D.~M.,\ 2001, MNRAS, 323, L17
\bibitem[Harris \& Krawczynski(2006)]{Harris06}
Harris, D.~E., \& Krawczynski, H.,\ 2006, ARA\&A, 44, 463
\bibitem[Harris et al.(2000)]{Harris00}
Harris, D.~E., et al.,\ 2000, ApJ, 530, L81
\bibitem[]{}
Henri G., Pelletier G., Roland J., 1993, ApJ, 404, L41 
\bibitem[Hirotani et al.(1999)]{Hirotani99}
Hirotani, K., Iguchi, S., Kimura, M., \& Wajima, K.,\ 1999, PASJ, 51, 263
\bibitem[Hirotani et al.(2000)]{Hirotani00}
Hirotani, K., Iguchi, S., Kimura, M.,
\& Wajima, K.,\ 2000, ApJ, 545, 100
\bibitem[Hirotani(2005)]{Hirotani05}
Hirotani, K.,\ 2005, ApJ, 619, 73
\bibitem[Homan et al.(2009)]{Homan09}
Homan, D.~C., Lister, M.~L., Aller, H.~D.,
Aller, M.~F., \& Wardle, J.~F.~C.,\ 2009,
ApJ, 696, 328
\bibitem[Hoshino et al.(1992)]{Hoshino92} Hoshino, M., Arons, J.,
Gallant, Y.~A., \& Langdon, A.~B.\ 1992, ApJ, 390, 454
\bibitem[]{}
Inoue S., Takahara F., 1996, ApJ, 463, 555
\bibitem[Isobe et al.(2002)]{Isobe02}
Isobe, N., Tashiro, M., Makishima, K., Iyomoto, N., Suzuki, M.,
Murakami, M.~M., Mori, M., \& Abe, K.,\ 2002, ApJ, 580, L111
\bibitem[Isobe et al.(2005)]{Isobe05}
Isobe, N., Makishima, K., Tashiro, M., \& Hong, S.,\ 2005, ApJ, 632, 781
\bibitem[]{}
Isobe N., Seta H., Gandhi P., Tashiro M.~S., 2011, ApJ, 727, 82 
\bibitem[]{}
Isobe N., Koyama S., 2015, PASJ, 67, 77 
\bibitem[Ito et al.(2008)]{Ito08} Ito, H., Kino, M., Kawakatu,
N., Isobe, N., \& Yamada, S.,\ 2008, ApJ, 685, 828 (I08)
\bibitem[Iwamoto \& Takahara(2002)]{Iwamoto02}
Iwamoto, S., \& Takahara, F.,\ 2002, ApJ,
565, 163
\bibitem[Iwamoto
\& Takahara(2004)]{Iwamoto2004}
Iwamoto, S., \& Takahara, F.\ 2004, ApJ, 601, 78
\bibitem[]{}
Kaiser C.~R., Schoenmakers A.~P., R{\"o}ttgering H.~J.~A., 2000, MNRAS, 315, 381
\bibitem[]{}
Kaiser C.~R., Alexander P., 1997, MNRAS, 286, 215
\bibitem[]{}
Kang S., Chen L., Wu Q., 2014, arXiv, arXiv:1409.3233
\bibitem[]{}
Kataoka J., Stawarz {\L}., 2005, ApJ, 622, 797
\bibitem[Kataoka et al.(2008)]{Kataoka08}
Kataoka, J., et al.,\ 2008, ApJ, 672, 787
\bibitem[Kawakatu \& Kino(2006)]{Kawakatu06}
Kawakatu, N., \& Kino, M.,\ 2006, MNRAS, 370,
1513
\bibitem[]{}
Kawakatu N., Nagai H., Kino M., 2008, ApJ, 687, 141 
\bibitem[Kellermann et al.(2004)]{Kellermann04}
Kellermann, K.~I., et al.,\ 2004, ApJ, 609, 539
\bibitem[Kino \& Takahara (2004)]{Kino04}
Kino, M., Takahara, F.,\ 2004, MNRAS,
349, 336
\bibitem[Kino \& Kawakatu(2005)]{Kino05}
Kino, M., \& Kawakatu, N.,\ 2005, MNRAS,
364, 659
\bibitem[Kino \& Takahara(2008)]{Kino08}
Kino, M., \& Takahara, F.,\ 2008, MNRAS, 383, 713
\bibitem[]{}
Kino M., Kawakatu N., Takahara F., 2012, ApJ, 751, 101 (KKT12) 
\bibitem[Kirk et al.(2000)]{Kirk00} Kirk, J.~G., Guthmann,
A.~W., Gallant, Y.~A., \& Achterberg, A.\ 2000, ApJ, 542, 235
\bibitem[Komissarov et al.(2007)]{Komissarov07} Komissarov, S.~S.,
Barkov, M.~V., Vlahakis, N., K{\"o}nigl, A.\ 2007, MNRAS, 380, 51
\bibitem[]{}
Konigl A., 1989, ApJ, 342, 208
\bibitem[Krichbaum et al.(1998)]{Krichbaum98}
Krichbaum, T.~P., Alef, W.,Witzel, A., Zensus, J.~A.,
Booth, R.~S., Greve, A., \& Rogers, A.~E.~E.,\ 1998, A\&A, 329, 873
\bibitem[]{}
Laing R.~A., Canvin J.~R., Bridle A.~H., Hardcastle M.~J., 2006, MNRAS, 
372, 510 
\bibitem[]{}
Laing R.~A., Bridle A.~H., 2002, MNRAS, 336, 1161 
\bibitem[Lazio et al.(2006)]{Lazio06}
Lazio, T.~J.~W., Cohen, A.~S., Kassim, N.~E.,
Perley, R.~A., Erickson, W.~C.,
Carilli, C.~L., \& Crane, P.~C.,\ 2006, ApJ, 642, L33
\bibitem[Lister et al.(2001)]{Lister01}
Lister, M.~L., Tingay, S.~J., Murphy, D.~W.,
Piner, B.~G., Jones, D.~L., \& Preston, R.~A.,\ 2001, ApJ, 554, 948
\bibitem[]{}
Machalski J., Jamrozy M., Konar C., 2010, A\&A, 510, A84
\bibitem[]{}
Maciel T., Alexander P., 2014, MNRAS, 442, 3469
\bibitem[McKinney(2006)]{McKinney06}
McKinney, J.~C.,\ 2006, MNRAS, 368, 1561
\bibitem[]{}
McKinney J.~C., Tchekhovskoy A., Blandford R.~D., 2012, MNRAS, 423, 308
\bibitem[]{}
McKinney J.~C., Tchekhovskoy A., Sadowski A., Narayan R., 2014, MNRAS, 441, 3177 
\bibitem[Mehta et al.(2009)]{Mehta09}
Mehta, K.~T., Georganopoulos, M., Perlman, E.~S.,
Padgett, C.~A., \& Chartas, G.,\ 2009, ApJ, 690, 1706
\bibitem[]{}
Meier D.~L., 2003, NewAR, 47, 667
\bibitem[]{}
Meisenheimer K., Yates M.~G., Roeser H.-J., 1997, A\&A, 325, 57
\bibitem[Mizuta et al.(2004)]{Mizuta04}
Mizuta, A., Yamada, S., \& Takabe, H.,\ 2004, ApJ, 606, 804
\bibitem[]{}
Mizuta A., Kino M., Nagakura H., 2010, ApJ, 709, L83
\bibitem[]{}
Moderski R., Sikora M., Madejski G.~M., Kamae T., 2004, ApJ, 611, 770 
\bibitem[]{}
Nakamura M., Tregillis I.~L., Li H., Li S., 2008, ApJ, 686, 843 
\bibitem[]{}
O'Sullivan S.~P., et al., 2013, ApJ, 764, 162 
\bibitem[]{}
Park J., Caprioli D., Spitkovsky A., 2014, arXiv, arXiv:1412.0672 
\bibitem[Perucho \& Mart{\'{\i}}]{Perucho07}
Perucho, M., \& Mart{\'{\i}} J.~M.,\ 2007, MNRAS, 382, 526
\bibitem[]{}
Phinney E.~S., 1982, MNRAS, 198, 1109
\bibitem[]{} 
Rees M.~J., 1984, ARA\&A, 22, 471
\bibitem[]{} 
Reynolds C.~S., Heinz S., Begelman M.~C., 2002, MNRAS, 332, 271 
\bibitem[Ruszkowski \& Begelman(2002)]{Ruszkowski02}
Ruszkowski, M., \& Begelman, M.~C.,\ 2002,
ApJ, 573, 485
\bibitem[]{}
Seta H., Tashiro M.~S., Inoue S., 2013, PASJ, 65, 106 
\bibitem[Scheck et al.(2002)]{Scheck02}
Scheck, L., Aloy, M.~A., Mart{\'{\i}}, J.~M.,
G{\'o}mez, J.~L., \& M{\"u}ller, E.,\ 2002, MNRAS, 331, 615
\bibitem[]{}
Sikora M., Sol H., Begelman M.~C., Madejski G.~M., 1996, MNRAS, 280, 781
\bibitem[Sikora \& Madejski(2000)]{Sikora00}
Sikora, M., \& Madejski, G.,\ 2000, ApJ,  534, 109
\bibitem[]{} 
Sikora M., Janiak M., Nalewajko K., Madejski G.~M., Moderski R., 2013, ApJ, 
779, 68 
\bibitem[Sironi \& Spitkovsky(2011)]{Sironi11}
Sironi, L., \& Spitkovsky, A.,\ 2011, ApJ, 726, 75
\bibitem[]{}
Sol H., Pelletier G., Asseo E., 1989, MNRAS, 237, 411
\bibitem[Spitkovsky(2008)]{Spitkovsky08}
Spitkovsky, A.,\ 2008, ApJ, 682, L5
\bibitem[Stawarz et al.(2007)]{Stawarz07}
Stawarz, L., Cheung, C.~C., Harris, D.~E.,
\& Ostrowski, M.,\ 2007, ApJ, 662, 213
\bibitem[]{}
Stawarz {\L}., et al., 2013, ApJ, 766, 48
\bibitem[Svensson(1982)]{Svensson82}
Svensson, R.,\ 1982, ApJ, 258, 335
\bibitem[Tashiro et al.(1998)]{Tashiro98}
Tashiro, M., et al.,\ 1998, ApJ, 499, 713
\bibitem[Tashiro et al.(2010)]{Tashiro10}
Tashiro, M.~S., Isobe, N., Seta, H.,
Matsuta, K., \& Yaji, Y.,\ 2010, PASJ, 61, 327
\bibitem[]{}
Tavecchio F., Maraschi L., Sambruna R.~M., 
Urry C.~M., 2000, ApJ, 544, L23 
\bibitem[]{}
Tchekhovskoy A., Narayan R., McKinney J.~C., 2011, MNRAS, 418, L79
\bibitem[]{}
Toma K., Takahara F., 2013, PTEP, 2013, 083E02
\bibitem[Uchiyama et al.(2005)]{Uchiyama05}
Uchiyama, Y., Urry, C.~M., Van Duyne, J.,
Cheung, C.~C., Sambruna, R.~M., Takahashi, T., Tavecchio, F.,
\& Maraschi, L.,\ 2005, ApJ, 631, L113
\bibitem[Wardle et al.(1998)]{Wardle98}
Wardle, J.~F.~C., Homan, D.~C., Ojha, R.,
\& Roberts, D.~H.,\ 1998, Nature, 395, 457
\bibitem[Watanabe et al.(2009)]{Watanabe09}
Watanabe, S., et al.\ 2009, ApJ, 694, 294
\bibitem[Wikson et al.(2000)]{Wilson00} Wilson, A.~S., Young, A.~J.,
\& Shopbell, P.~L.,\ 2000, ApJ, 544, L27
\bibitem[Wilson et al.(2006)]{Wilson06}
Wilson, A.~S., Smith, D.~A., \& Young, A.~J.,\ 2006, ApJ, 644, L9
\bibitem[Yaji et al.(2010)]{Yaji10} Yaji, Y., Tashiro, M., Isobe, N., 
Kino, M., Asada, K., Nagai, H., Koyama, S., \& Kusunose, M.,\ 2010, ApJ, 714, 37
\bibitem[Yamasaki et al.(1999)]{Yamasaki99}
Yamasaki, T., Takahara, F., \& Kusunose, M.,\ 1999, ApJ, 523, L21
\end{thebibliography}
\end{document}